\pdfoutput=1
\newcommand*{\ATLASLATEXPATH}{}
\documentclass[UKenglish,texlive=2016,cernpreprint]{\ATLASLATEXPATH atlasdoc}
\usepackage[subcaption, biblatex=true]{\ATLASLATEXPATH atlaspackage}
\usepackage[block=space]{\ATLASLATEXPATH atlasbiblatex}
\usepackage{enumitem}
\usepackage{\ATLASLATEXPATH atlascontribute}

\usepackage{\ATLASLATEXPATH atlasphysics}
\usepackage{\ATLASLATEXPATH atlasmisc}

\usepackage{graphicx}
\usepackage{epstopdf}
\usepackage{multirow}
\usepackage{booktabs}
\usepackage{siunitx}

\let\sc\textsc

\let\rm\mathrm
\let\bf\textbf

\graphicspath{{logos/}{}}

\usepackage{SUSY-2016-24-PAPER-defs}

\AtlasTitle{Search for electroweak production of supersymmetric particles in final states with two or three leptons at $\boldmath{\sqrt{s}=13\,}$TeV with the ATLAS detector}
\author{The ATLAS Collaboration}
\usepackage{authblk}

 \PreprintIdNumber{CERN-EP-2017-303}

 \AtlasNote{SUSY-2016-24}

\AtlasJournal{EPJC}
 \AtlasJournalRef{Eur. Phys. J. C 78 (2018) 995}
 \AtlasDOI{10.1140/epjc/s10052-018-6423-7}

\AtlasAbstract{%

A search for the electroweak production of charginos, neutralinos and sleptons decaying into final states involving two or three electrons or muons is presented. 
The analysis is based on 36.1~fb$^{-1}$ of $\sqrt{s}=13$~TeV proton--proton collisions recorded by the ATLAS detector at the Large Hadron Collider. 
Several scenarios based on simplified models are considered. These include the associated production of the next-to-lightest neutralino and the lightest chargino, 
followed by their decays into final states with leptons
and the lightest neutralino via
either sleptons or Standard Model gauge bosons; direct production of chargino pairs, which in turn decay into leptons and the lightest neutralino 
via intermediate sleptons; 
and slepton pair production,
where each slepton decays directly into the lightest neutralino and a lepton. 
No significant deviations from the Standard Model expectation are observed and stringent limits at 95\% confidence level are placed
on the masses of relevant supersymmetric particles in each of these scenarios. For a massless lightest neutralino, masses up to 580~GeV are excluded for the associated production of the next-to-lightest neutralino and the lightest chargino, assuming gauge-boson mediated decays, whereas for slepton-pair production masses up to 500 GeV are excluded assuming three generations of mass-degenerate sleptons.
 }

\hypersetup{pdftitle={ATLAS document},pdfauthor={The ATLAS Collaboration}}
\begin{document}

\maketitle

\section{Introduction}
\label{sec:intro}
Supersymmetry (SUSY)~\cite{Golfand:1971iw,Volkov:1973ix,Wess:1974tw,Wess:1974jb,Ferrara:1974pu,Salam:1974ig,Martin:1997ns} is one of the most studied extensions 
of the Standard Model (SM).
In its minimal realization (the Minimal Supersymmetric Standard Model, or MSSM)~\cite{Fayet:1976et,Fayet:1977yc},
it predicts new fermionic (bosonic) partners of the fundamental SM bosons (fermions)
and an additional Higgs doublet. These new SUSY particles, or sparticles,
can provide an elegant solution to the gauge hierarchy problem~\cite{Sakai:1981gr,Dimopoulos:1981yj,Ibanez:1981yh,Dimopoulos:1981zb}. 
In $R$-parity-conserving models~\cite{Farrar:1978xj}, sparticles can only be produced in pairs and 
the lightest supersymmetric particle (LSP) is stable. This is typically assumed to be the \ninoone 
neutralino\footnote{The SUSY partners of the Higgs field (known as higgsinos) and of the electroweak gauge fields (the bino for the U(1) gauge field and winos for the $W$ fields) mix to form the mass eigenstates known as charginos and neutralinos.},  which can then provide a natural candidate for dark matter~\cite{Goldberg:1983nd,Ellis:1983ew}. 
If produced in proton--proton collisions, a neutralino LSP would escape detection
and lead to an excess of events with large missing transverse momentum above the expectations from SM processes,
a characteristic that is exploited to search for SUSY signals in analyses presented in this paper.

The production cross-sections of SUSY particles at the Large Hadron Collider (LHC)~\cite{1748-0221-3-08-S08001} depend both on the type of interaction involved 
and on the masses of the sparticles. 
The coloured sparticles (squarks and gluinos) are produced in strong interactions with significantly larger production cross-sections 
than non-coloured sparticles of equal masses, such as the charginos ($\tilde{\chi}^{\pm}_{i}$, $i = 1, 2$) and neutralinos ($\tilde{\chi}^{0}_{j}$, $j = 1, 2, 3, 4$) and the
sleptons ($\slepton$ and $\snu$). The direct production of charginos and neutralinos or slepton pairs can dominate SUSY production at the LHC if the masses of the gluinos and the squarks are significantly larger. 
With searches performed by the ATLAS~\cite{PERF-2007-01} and CMS~\cite{CMS-TDR-08-001} experiments 
during LHC Run 2, the exclusion limits on coloured sparticle masses extend up to approximately $2\,$TeV~\cite{Aaboud:2017vwy,Sirunyan:2017cwe,Sirunyan:2017kqq}, making electroweak production an increasingly important probe for SUSY signals at the LHC.

This paper presents a set of searches for the electroweak production of charginos, neutralinos and sleptons decaying into final states with two or 
three  electrons or muons using 36.1 fb$^{-1}$ of  proton--proton collision data delivered by the LHC at a centre-of-mass energy of $\sqrt{s}=13$~TeV. 
The results build on studies performed during LHC Run 1 at $\sqrt{s}=7$~TeV and 8~TeV by the ATLAS Collaboration~\cite{SUSY-2013-11,SUSY-2013-12,Aad:2015eda}.
Analogous studies by the CMS Collaboration are presented in Refs.~\cite{Sirunyan:2017lae,CMS-SUS-12-006,Khachatryan:2014mma,Khachatryan:2014qwa}. 

After descriptions of the SUSY scenarios considered (Section~\ref{sec:signal}), the experimental apparatus (Section~\ref{sec:detector}), 
the simulated samples~(Section~\ref{sec:dataMCsamples}) and the event reconstruction (Section~\ref{sec:objects}), the analysis search strategy is discussed in Section~\ref{sec:analysis}. 
This is followed by Section~\ref{sec:backgrounds}, which describes the estimation of 
SM contributions to the measured yields in the signal regions, and by Section~\ref{sec:systematics}, which 
discusses systematic uncertainties affecting the searches. Results are presented in Section~\ref{sec:result}, together with 
the statistical tests used to interpret them in the context of relevant SUSY benchmark scenarios. 
Section~\ref{sec:conclusion} summarizes the main conclusions.

\section{SUSY scenarios and search strategy}
\label{sec:signal}
\begin{figure}[h]
   \centering

\begin{subfigure}[t]{0.3\textwidth}\includegraphics[width=\textwidth]{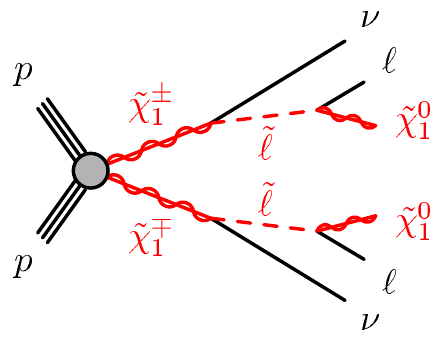}\caption{}\label{fig:C1C1tree}\end{subfigure}
\begin{subfigure}[t]{0.3\textwidth}\includegraphics[width=\textwidth]{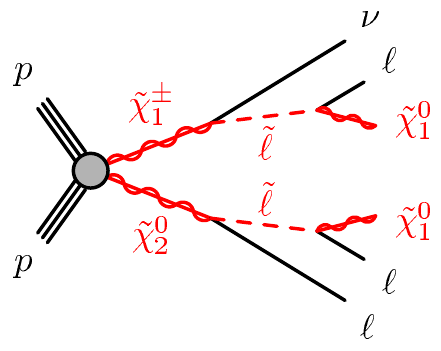}\caption{}\label{fig:C1N2tree}\end{subfigure} \\
\begin{subfigure}[t]{0.3\textwidth}\includegraphics[width=\textwidth]{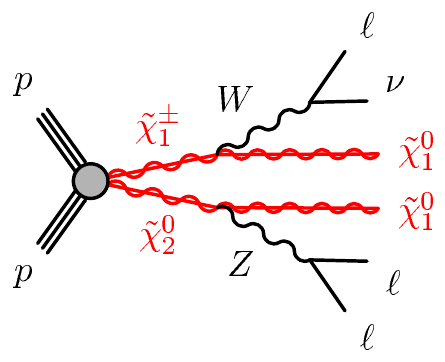}\caption{}\label{fig:C1N2treeWZ}\end{subfigure}
\begin{subfigure}[t]{0.3\textwidth}\includegraphics[width=\textwidth]{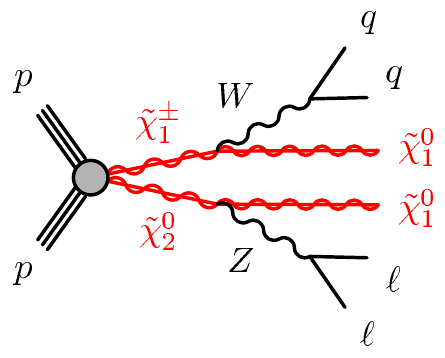}\caption{}\label{fig:C1N2treeWZj}\end{subfigure}
\begin{subfigure}[t]{0.3\textwidth}\includegraphics[width=\textwidth]{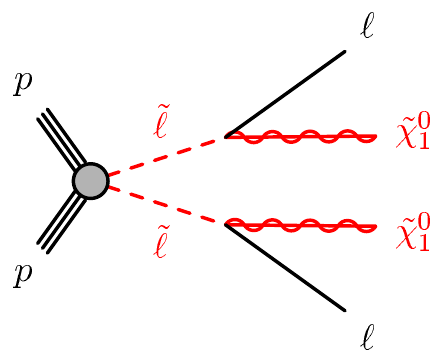}\caption{}\label{fig:sleptontree}\end{subfigure} \\
\caption{Diagrams of physics scenarios studied in this paper:
(a) $\chinoonep\chinoonem$ production with $\slepton$-mediated  decays into final states with two leptons,
(b) $\chinoonepm\ninotwo$ production with $\slepton$-mediated decays into final states with three leptons,
(c) $\chinoonepm\ninotwo$ production with decays via leptonically decaying $W$ and $Z$ bosons into final states with three leptons, 
(d) $\chinoonepm\ninotwo$ production with decays via a hadronically decaying $W$ boson and a leptonically decaying $Z$ boson into final states with two leptons and two jets, and
(e) slepton pair production with decays into final states with two leptons.
}
   \label{fig:TreeDiagramsPhysScenarios}
 \end{figure}

This paper presents searches for the direct pair-production of $\chinoonep\chinoonem$, $\chinoonepm\ninotwo$ and 
$\slepton\slepton$ particles, in final states with exactly two or 
three electrons and muons, two $\ninoone$  particles, and possibly additional jets or neutrinos. 
Simplified models~\cite{Alwall:2008ag}, in which the masses of the relevant sparticles are the only free parameters, 
are used for interpretation and to guide the design of the searches.
The pure wino $\chinoonepm$ and $\ninotwo$ are taken to be 
mass-degenerate, and so are the scalar partners of the left-handed charged leptons and neutrinos ($\tilde{e}_{\textup{L}}$, $\tilde{\mu}_{\textup{L}}$, $\tilde{\tau}_{\textup{L}}$ and $\tilde{\nu}$). 
Intermediate slepton masses, when relevant, are chosen to be midway between the mass of the heavier chargino and neutralino and that of the lightest neutralino, which is pure bino, 
and equal branching ratios for the three slepton flavours are assumed. The analysis sensitivity is not expected to depend strongly on the slepton mass hypothesis for a broad range 
of slepton masses, while it degrades as the slepton mass approaches that of the heavier chargino and neutralino, leading to lower $p_T$ values for the leptons produced in the heavy 
chargino and neutralino decays~\cite{Aad:2015eda}.
Lepton flavour is conserved in all models.
Diagrams of processes considered are shown in Figure~\ref{fig:TreeDiagramsPhysScenarios}. 
For models exploring $\chinoonep\chinoonem$ production, it is assumed that the sleptons are also light and thus accessible in the sparticle decay chains, as illustrated in Figure~\ref{fig:C1C1tree}. 
Two different classes of models are considered for $\chinoonepm\ninotwo$ production: in one case, the $\chinoonepm$ chargino and $\ninotwo$ neutralino can decay into final-state SM particles and a $\ninoone$ neutralino via 
an intermediate $\slepton_{\textup{L}}$ or $\tilde{\nu}_{\textup{L}}$, with a branching ratio of 50\% to each (Figure~\ref{fig:C1N2tree}); in the other case the $\chinoonepm$ chargino and $\ninotwo$ neutralino decays 
proceed via SM gauge bosons ($W$ or $Z$). 
For the gauge-boson-mediated decays, two distinct final states are considered: three-lepton (where lepton refers to an electron or muon) events where both
the $W$ and $Z$ bosons decay leptonically (Figure~\ref{fig:C1N2treeWZ}) or events with two opposite-sign leptons and two jets where the
$W$ boson decays hadronically and the $Z$ boson decays leptonically (Figure~\ref{fig:C1N2treeWZj}).
In models with direct $\slepton\slepton$ production, each 
slepton decays into a lepton and a $\ninoone$ with a 100\% branching ratio (Figure~\ref{fig:sleptontree}), and $\tilde{e}_{\textup{L}}$, $\tilde{e}_{\textup{R}}$, $\tilde{\mu}_{\textup{L}}$, $\tilde{\mu}_{\textup{R}}$, $\tilde{\tau}_{\textup{L}}$ and $\tilde{\tau}_{\textup{R}}$ are assumed to be mass-degenerate.

Events are recorded using triggers requiring the presence of at least two leptons and assigned to one of three mutually exclusive analysis channels depending on the
lepton and jet multiplicity. The 2$\ell$+0jets channel targets chargino- and slepton-pair production,
the 2$\ell$+jets channel targets chargino-neutralino production with gauge-boson-mediated decays,
and the 3$\ell$ channel targets chargino-neutralino production with slepton- or gauge-boson-mediated decays.
For each channel, a set of signal regions (SR), defined in Section~\ref{sec:analysis}, use requirements on \MET and other kinematic quantities, which are
optimized for different SUSY models and sparticle masses. The analyses employ ``inclusive'' SRs to quantify significance without assuming a particular
signal model and to exclude regions of SUSY model parameter space, as well as sets of orthogonal ``exclusive'' SRs that are considered simultaneously
during limit-setting to improve the exclusion sensitivity.

\section{ATLAS detector}
\label{sec:detector}
The ATLAS experiment~\cite{PERF-2007-01} is a multi-purpose particle detector with a forward-backward symmetric cylindrical
geometry and nearly $4\pi$ coverage in solid angle.\footnote{ATLAS uses
  a right-handed coordinate system with its origin at the nominal
  interaction point (IP) in the centre of the detector and the
  $z$-axis along the beam direction. The $x$-axis points from the IP to the
  centre of the LHC ring, and the $y$-axis points upward. Cylindrical
  coordinates ($r$, $\phi$) are used in the transverse plane, $\phi$
  being the azimuthal angle around the beam direction. The pseudorapidity
  is defined in terms of the polar angle $\theta$ as $\eta = -\ln
  \tan(\theta/2)$.
  Angular distance is measured in units of $\Delta R \equiv \sqrt{ (\Delta \eta)^2 + (\Delta \phi)^2 }$.
  The transverse momentum, \pt, and energy, \et, are defined with respect to the beam axis ($x$--$y$ plane).
  }
The interaction point is surrounded by an inner detector (ID), a
calorimeter system, and a muon spectrometer.

The ID provides precision tracking of charged particles for
pseudorapidities $|\eta| < 2.5$ and is surrounded by a superconducting solenoid providing a \SI{2}{T} axial magnetic field.
The ID consists of silicon pixel and microstrip detectors inside a
transition radiation tracker.  
One significant upgrade for the $\sqrt{s}=13$~TeV running period is the installation of the
insertable B-layer~\cite{ATLAS-TDR-19}, an additional pixel layer close to the interaction point which provides high-resolution hits at small radius to improve the tracking performance.

In the pseudorapidity region $|\eta| < 3.2$, high-granularity lead/liquid-argon (LAr)
electromagnetic (EM) sampling calorimeters are used.
A steel/scintillator tile calorimeter measures hadron energies for
$|\eta| < 1.7$.
The endcap and forward regions, spanning $1.5<|\eta| <4.9$, are 
instrumented with LAr calorimeters, for both the EM and hadronic
measurements.

The muon spectrometer consists of three large superconducting toroids
with eight coils each, 
and a system of trigger and precision-tracking chambers, 
which provide triggering and tracking capabilities in the
ranges $|\eta| < 2.4$ and $|\eta| < 2.7$, respectively.

A two-level trigger system is used to select events \cite{Aaboud:2016leb}. The first-level
trigger is implemented in hardware and uses a subset of the detector
information. This is followed by the software-based high-level trigger,
which runs offline reconstruction and
calibration software, reducing the event rate to about \SI{1}{kHz}.

\section{Data and simulated event samples}
\label{sec:dataMCsamples}
This analysis uses proton--proton collision data delivered by the LHC at $\sqrt{s}=13\TeV$ in 2015 and 2016.
After fulfilling data-quality requirements, the data sample amounts to an integrated luminosity of 36.1~fb$^{-1}$. This value is derived using a methodology similar to that detailed in Refs.~\cite{DAPR-2013-01}, from a calibration of the luminosity scale using $x$--$y$ beam-separation scans performed in August 2015 and May 2016.

Various samples of Monte Carlo (MC) simulated events are used to model the SUSY signal and help
in the estimation of the SM backgrounds. The samples include an ATLAS detector simulation~\cite{SOFT-2010-01}, based on {\sc Geant4}~\cite{Agostinelli:2002hh}, or a fast simulation~\cite{SOFT-2010-01} that uses a parameterization of the calorimeter response~\cite{ATL-PHYS-PUB-2010-013} and {\sc Geant4} for the other parts of the detector. The simulated events are reconstructed in the same manner as the data. 

Diboson processes were simulated with the \SHERPA~v2.2.1 event
generator~\cite{Gleisberg:2008ta, ATL-PHYS-PUB-2016-002}
and normalized using next-to-leading-order (NLO) cross-sections~\cite{diboson1,diboson2}. The matrix
elements containing all diagrams with four electroweak vertices with
additional hard parton emissions were calculated with {\sc Comix}~\cite{Gleisberg:2008fv}  and
virtual QCD corrections were calculated with {\sc OpenLoops}~\cite{Cascioli:2011va}.  Matrix element
calculations were merged with the \SHERPA parton shower~\cite{Schumann:2007mg} using the
ME+PS@NLO prescription~\cite{Hoeche:2012yf}. The NNPDF3.0 NNLO parton distribution function (PDF) set~\cite{Ball:2014uwa} was used in
conjunction with dedicated parton shower tuning developed by the
\SHERPA authors. 
The fully leptonic channels were
calculated at NLO in the strong coupling constant with up to one additional parton for $4\ell$ and
$2\ell+2\nu$, at NLO with no additional parton for $3\ell+\nu$, and at leading order (LO) with up to three additional partons. 
Processes with one of the
bosons decaying hadronically and the other leptonically were calculated
with up to one additional parton at NLO and up to three additional
partons at LO. 

Diboson processes with six electroweak vertices, such as
same-sign $W$ boson production in association with two
jets, $W^\pm W^\pm jj$, and triboson processes were simulated as above 
with \SHERPA~v2.2.1 using the NNPDF3.0 PDF set. Diboson
processes with six vertices were calculated at LO with up to one additional parton.
Fully leptonic triboson processes ($WWW$, $WWZ$, $WZZ$ and $ZZZ$) were
calculated at LO with up to two additional partons and at NLO for the
inclusive processes and normalized using NLO cross-sections.

Events containing \Zboson bosons and associated jets ($Z/\gamma^*$+jets, also referred to as $Z$+jets in the following) were also produced using the  \SHERPA v2.2.1 generator with massive $b/c$-quarks to improve the treatment of the associated production of $\Zboson$ bosons with jets containing $b$- and $c$-hadrons~\cite{ATL-PHYS-PUB-2016-003}. Matrix elements were calculated with up to two additional partons at NLO and up to four additional partons at LO, using {\sc Comix}, {\sc OpenLoops},  and \SHERPA parton shower with  ME+PS@NLO in a way similar to that described above.  A global $K$-factor was used to normalize the $Z$+jets events to the next-to-next-to-leading-order (NNLO) QCD cross-sections~\cite{Catani:2009sm}.

For the production of \ttbar and single top quarks in the $Wt$ channel, the \POWHEGBOX v2~\cite{Re:2010bp,Frixione:2007nw} generator with the CT10 PDF set~\cite{CT10pdf} was used, as discussed in Ref.~\cite{ATL-PHYS-PUB-2016-004}. 
The top quark mass was set at 172.5~GeV for all MC samples involving top quark production. The \ttbar events were normalized using the NNLO+next-to-next-to-leading-logarithm (NNLL) QCD~\cite{Czakon:2011xx} cross-section, while the cross-section for single-top-quark 
events was calculated at NLO+NNLL~\cite{Kidonakis:2010ux}.

Samples of $t\bar{t} V$ (with $V=W$ and $Z$, including non-resonant $Z/\gamma^*$ contributions) 
and $t\bar{t}WW$ production were generated at LO with \MGMCatNLO v2.2.2~\cite{Alwall:2014hca} interfaced to \PYTHIA 8.186~\cite{Sjostrand:2007gs} for parton showering, hadronisation and the description of the underlying event, with up to two ($\ttbar W$), one ($\ttbar Z$) or no ($\ttbar WW$) extra partons included in the matrix element, 
as described in Ref.~\cite{ATL-PHYS-PUB-2016-005}. 
\MADGRAPH was also used to simulate the $tZ$, $\ttbar\ttbar$ and $\ttbar t$ processes. 
A set of tuned parameters called the {\sc A14} tune~\cite{ATL-PHYS-PUB-2014-021} was used together with the {\sc NNPDF2.3LO} PDF set~\cite{Ball:2012cx}. 
The $\ttbar W$, $\ttbar Z$, $\ttbar WW$ and $\ttbar\ttbar$ events were normalized using their NLO cross-section~\cite{ATL-PHYS-PUB-2016-005} 
while the generator cross-section was used for $tZ$ and $\ttbar t$.

Higgs boson production processes (including gluon--gluon fusion, associated $VH$ production and vector-boson fusion) were generated
using \POWHEGBOX v2~\cite{Alioli:2010xd} and \PYTHIA 8.186 and normalized using cross-sections calculated at NNLO
with soft gluon emission effects added at NNLL accuracy~\cite{deFlorian:2016spz}, whilst $\ttbar H$
events were produced using \MGMCatNLO 2.3.2 + \Herwigpp~\cite{Bahr:2008pv} and normalized using the NLO cross-section~\cite{ATL-PHYS-PUB-2016-005}.
All samples assume a Higgs boson mass of 125 GeV.

The SUSY signal processes were generated from LO matrix elements with up to two extra partons, 
using the \MADGRAPH v2.2.3 generator interfaced to \PYTHIA 8.186 with the {\sc A14} tune for the modelling of the SUSY decay chain, 
parton showering, hadronization and the description of the underlying event. 
Parton luminosities were provided by the {\sc NNPDF2.3LO} PDF set. 
Jet--parton matching was realized following the CKKW-L prescription~\cite{Lonnblad:2011xx}, with a matching scale set to one quarter of the pair-produced superpartner mass. 
Signal cross-sections were calculated at NLO,
with soft gluon emission effects added at next-to-leading-logarithm (NLL) accuracy~\cite{Beenakker:1996ch,Kulesza:2008jb,Kulesza:2009kq,Beenakker:2009ha,Beenakker:2011fu}. 
The nominal cross-section and its uncertainty were taken from an envelope of cross-section predictions using different PDF sets and factorization and renormalization scales, 
as described in Ref.~\cite{Borschensky:2014cia}.
The cross-section for $\chinoonep\chinoonem$ production, each with a mass of $600$~GeV, is $9.50\pm0.91$~fb,
while the cross-section for $\chinoonepm\ninotwo$ production, each with a mass of $800$~GeV, is $4.76\pm0.56$~fb.

In all MC samples, except those produced by \SHERPA, the {\sc EvtGen}~v1.2.0 program~\cite{Lange:2001uf} was used to model the properties of $b$- and $c$-hadron decays. 
To simulate the effects of additional $pp$ collisions per bunch crossing (pile-up), 
additional interactions were generated using the soft QCD processes of \PYTHIA 8.186 
with the A2 tune~\cite{ATLAS-PHYS-PUB-2012-003} and the MSTW2008LO PDF set~\cite{Martin:2009iq}, 
and overlaid onto the simulated hard-scatter event. 
The Monte Carlo samples were reweighted so that the distribution of the number of pile-up interactions matches the distribution in data.

\section{Event reconstruction and preselection}
\label{sec:objects}
 Events used in the analysis were recorded during stable data-taking conditions and must have a reconstructed primary vertex~\cite{ATL-PHYS-PUB-2015-026} with at least two 
associated tracks with $\pt>400~\MeV$. The primary vertex of an event is identified as the vertex with the highest $\Sigma \pt^2$ of associated tracks.

Two identification criteria are defined for the objects used in these analyses, referred to as ``baseline'' and ``signal'' (with the signal objects being a subset of the baseline ones).
The former are defined to disambiguate between overlapping physics objects and to perform data-driven estimations of non-prompt leptonic backgrounds (discussed in Section~\ref{sec:backgrounds})
while the latter are used to construct kinematic and multiplicity discriminating variables needed for the event selection.

Baseline electrons are reconstructed from isolated electromagnetic 
calorimeter energy deposits matched to ID tracks and are required to have $|\eta|<2.47$, 
$\pT>\SI{10}{\GeV}$, 
and to pass a loose likelihood-based identification requirement~\cite{ATL-PHYS-PUB-2015-041,PERF-2016-01}. 
The likelihood input variables include measurements of calorimeter shower shapes and track properties from the ID.

Baseline muons are reconstructed in the region $|\eta|<2.7$ 
from muon spectrometer tracks matching ID tracks.
All muons must have $\pT>\SI{10}{\GeV}$ and must pass the ``medium identification'' requirements defined in Ref.~\cite{PERF-2015-10}, 
based on selection of the number of hits
and curvature measurements in the ID and muon spectrometer systems.

Jets are reconstructed with the anti-$k_t$
algorithm~\cite{Cacciari:2008} as implemented in the FastJet package~\cite{Cacciari:2011ma}, with radius parameter $R=0.4$, using three-dimensional energy
clusters in the calorimeter~\cite{Aad:2016upy} as input. 
Baseline jets must have $\pT>\SI{20}{\GeV}$ and $|\eta|<4.5$ and signal jets have the tighter requirement of $|\eta|<2.4$.
Jet energies are calibrated as described in Refs.~\cite{ATL-PHYS-PUB-2015-015,Aaboud:2017jcu}.
In order to reduce the effects of pile-up,  jets with $\pt<\SI{60}{GeV}$ and $|\eta|<2.4$ must have 
a significant fraction of their associated tracks compatible with originating from the primary vertex, as defined by the jet vertex tagger~\cite{PERF-2014-03}.
Furthermore, for all jets the expected average energy contribution from pile-up is subtracted according to the jet area~\cite{Cacciari:2008gn,PERF-2014-03}. 
Events are discarded if they contain any jet that is judged by basic quality criteria to be detector noise or non-collision background.

Identification of jets containing $b$-hadrons ($b$-jets), so called $b$-tagging, is performed with the MV2c10 algorithm, 
a multivariate discriminant making use of track impact parameters 
and reconstructed secondary vertices~\cite{ATL-PHYS-PUB-2016-012,PERF-2012-04}.
A requirement is chosen corresponding to a 77\% average efficiency 
obtained for $b$-jets in simulated $\ttbar$ events. 
The corresponding rejection factors against jets originating from $c$-quarks, from $\tau$-leptons, and from light quarks and gluons in the same sample 
at this working point are 6, 22 and 134, respectively.

Baseline photon candidates are required to meet the ``tight'' selection criteria of Ref.~\cite{PERF-2013-04} and satisfy $\pt>25$~\GeV\ and $|\eta|<2.37$, but excluding the transition region $1.37<|\eta|<1.52$,
where the calorimeter performance is degraded.

After object identification, an ``object-removal procedure'' is performed on all baseline objects to remove possible double-counting in the reconstruction:
\vspace{-5pt}
\begin{enumerate}[nolistsep]
\item Any electron sharing an ID track with a muon is removed.
\item If a $b$-tagged jet (identified using the 85\% efficiency working point of the MV2c10 algorithm) is within $\Delta R = 0.2$ of an electron candidate, the electron is rejected, as it is likely to be from a semileptonic 
$b$-hadron decay; if the jet within $\Delta R = 0.2$ of the electron is not $b$-tagged, the jet itself is discarded, as it likely originates from an electron-induced shower. 
\item Electrons within $\Delta R=0.4$ of a remaining jet candidate are discarded, to further suppress electrons from semileptonic decays of $b$- and $c$-hadrons.
\item Jets with a nearby muon that carries a significant fraction of the transverse momentum of the jet 
($\pt^\mu > 0.7 \sum \pt^{\textrm{jet tracks}}$, where $\pt^\mu$ and $\pt^{\textrm{jet tracks}}$ are the transverse momenta of the muon and the tracks associated with the jet, respectively) are discarded either if the candidate muon is within $\Delta R=0.2$ of the jet or if the muon is matched to a track associated with the jet. Only jets with fewer than three associated tracks can be discarded in this step. 
\item Muons within $\Delta R=0.4$ of a remaining jet candidate are discarded to suppress muons from semileptonic decays of $b$- and $c$-hadrons.

\end{enumerate}

Signal electrons must satisfy a ``medium'' likelihood-based identification requirement~\cite{ATL-PHYS-PUB-2015-041} and the track associated with the electron must have a significance of the transverse impact parameter relative to the reconstructed primary vertex, $d_0$, of $\vert d_0\vert/\sigma(d_0) < 5$, with $\sigma(d_0)$ being the uncertainty in $d_0$. In addition, the longitudinal impact parameter (again relative to the reconstructed primary vertex), $z_0$, must satisfy $\vert z_0 \sin\theta\vert  < 0.5$~mm. Similarly, signal muons must satisfy the requirements of $\vert d_0\vert/\sigma(d_0) < 3$, $\vert z_0 \sin\theta\vert  < 0.5$~mm, and additionally have $|\eta|<2.4$.  Isolation requirements are also applied to both the signal electrons and muons to reduce the contributions of ``fake'' or non-prompt leptons, which originate from misidentified hadrons, photons conversions, and hadron decays. These $\pt$- and $\eta$-dependent requirements use track- and calorimeter-based information and have efficiencies in $Z\rightarrow\ee$ and $Z\rightarrow\mumu$ events that rise from 95\% at 25 GeV to 99\% at 60 GeV. 

The missing transverse momentum \pTmiss, with magnitude \MET, is the negative vector sum of the transverse momenta of 
all identified physics objects (electrons, photons, muons, jets) and an additional soft term. The soft term is constructed 
from all tracks that are not associated with any physics object and that are associated with the primary vertex, to 
suppress contributions from pile-up interactions. The \MET value is adjusted for the calibration of the jets and the 
other identified physics objects above~\cite{ATL-PHYS-PUB-2015-027}.

Events considered in the analysis must pass a trigger selection requiring either two electrons, two muons or an electron plus a muon.
The trigger-level thresholds on the \pt value of the leptons involved in the trigger decision are in the range 8--22 GeV and are looser than
those applied offline to ensure that trigger efficiencies are constant in the relevant phase space.

Events containing a photon and jets are used to estimate the $Z$+jets background in events with two leptons and jets. These events are selected with a set of prescaled single-photon
triggers with $\pt$ thresholds in the range 35--100~\GeV\ and an unprescaled single-photon trigger with threshold $\pt=140$ GeV.
Signal photons in this control sample must
have $\pt>37$~\GeV\ to be in the efficiency plateau of the lowest-threshold single-photon trigger,
fall outside the barrel-endcap transition region defined by $1.37 < |\eta| < 1.52$,
and pass ``tight'' selection criteria described in Ref.~\cite{PERF-2013-05}, as well as 
\pt- and $\eta$-dependent requirements on both track- and calorimeter-based isolation.

Simulated events are corrected to account for small differences in the signal lepton trigger, reconstruction, identification, isolation, as well as
$b$-tagging efficiencies between data and MC simulation.

\section{Signal regions}
\label{sec:analysis}
In order to search for the electroweak production of supersymmetric particles, three different search channels that target different SUSY processes are defined:

\begin{itemize}
\item \textbf{2$\ell$+0jets channel: } targets $\chinoonep\chinoonem$ and $\slepton\slepton$ production (shown in Figures~\ref{fig:C1C1tree} and~\ref{fig:sleptontree}) in signal regions with a jet veto and defined using the ``stransverse mass'' variable, $m_{\rm{T2}}$~\cite{Lester:1999tx,Barr:2003rg}, and the dilepton invariant mass \mll;
\item \textbf{2$\ell$+jets channel: } targets  $\chinoonepm\ninotwo$ production with decays via gauge bosons (shown in Figure~\ref{fig:C1N2treeWZj}) into two same-flavour opposite-sign (SFOS) leptons (from the $Z$ boson) and at least two jets (from the $W$ boson);
\item \textbf{3$\ell$ channel: } targets $\chinoonepm\ninotwo$ production with decays via intermediate $\slepton$ or gauge bosons into three-lepton final states (shown in Figures~\ref{fig:C1N2tree} and~\ref{fig:C1N2treeWZ}).
\end{itemize}

In each channel, inclusive and/or exclusive signal regions (SRs) are defined that require exactly two or three signal leptons, with vetos on any additional baseline leptons.
In the 2$\ell$+0jets channel only, this additional baseline lepton veto is applied before considering overlap-removal. The leading and sub-leading
leptons are required to have \pt$>25$ GeV and 20 GeV respectively; however, in the 2$\ell$+jets and 3$\ell$ channels, tighter lepton \pt\
requirements are applied to the sub-leading leptons.

\subsection{Signal regions for 2$\ell$+0jets channel}

In the 2$\ell$+0jets channel the leptons are required to be of opposite sign and events are separated into ``same flavour'' (SF) events (corresponding to dielectron, $e^{+}e^{-}$, and dimuon, $\mu^{+}\mu^{-}$, events) and ``different flavour'' (DF) events (electron--muon, $e^{\pm}\mu^{\mp}$). This division is driven by the different background compositions in the two classes of events. All events used in the SRs are required to have a dilepton invariant mass $m_{\ell\ell}>40$ GeV and
not contain any $b$-tagged jets with \pt$>20$ GeV or non-$b$-tagged jets with \pt$>60$~GeV.

After this preselection, exclusive signal regions are used to maximize exclusion sensitivity across the simplified model parameter space for $\chinoonep\chinoonem$ and $\slepton\slepton$ production.
In the SF regions a two-dimensional binning in \mttwo\ and $m_{\ell\ell}$ is used as high-$m_{\ell\ell}$ requirements provide strong suppression of the $Z$+jets background,
whereas in the DF regions, where the $Z$+jets background is negligible, a one-dimensional binning in \mttwo\ is sufficient. The stransverse mass \mttwo\ is defined as:
\[\mttwo = \min_{\qTvec}\left[\max\left(\mT(\pTell{1},\qTvec),\mT(\pTell{2},\pTmiss-\qTvec)\right)\right],\]
where $\pTell{1}$ and $\pTell{2}$ are the transverse momentum vectors of the two leptons,
and $\qTvec$ is a transverse momentum vector that minimizes the larger 
of $\mT(\pTell{1},\qTvec)
$ and $\mT(\pTell{2},\pTmiss-\qTvec)$, where:

\[
\mT(\pTvec,\qTvec) = \sqrt{2(\pT\qT-\pTvec\cdot\qTvec)}.
\]

For SM backgrounds of $\ttbar$ and $WW$ production in which the missing transverse momentum and the pair of selected leptons originate from two $W\to\ell\nu$ decays and all momenta are accurately measured,
the \mttwo\ value must be less than the $W$ boson mass $m_W$, and requiring the \mttwo\ value to significantly exceed $m_W$ thus strongly suppresses these backgrounds
while retaining high efficiency for many SUSY signals.

When producing model-dependent exclusion limits in the $\chinoonep\chinoonem$ simplified models, all signal regions are statistically combined, whereas only the 
same-flavour regions are used when probing $\slepton\slepton$ production. In addition, a set of inclusive signal regions are also defined, and these are used to provide a more model-independent test for an excess of events. The definitions of both the exclusive and inclusive signal regions are provided in Table~\ref{tab:exclusiveSR}.

\begin{table}[t]
   \centering
\caption{The definitions of the exclusive and inclusive signal regions for the 2$\ell$+0jets channel. Relevant kinematic variables are defined in the text.
The bins labelled ``DF''or ``SF'' refer to signal regions with different-flavour or same-flavour lepton pair combinations, respectively.
\label{tab:exclusiveSR}}

  \begin{tabular}{|c | c | c| c|}
\hline
\multicolumn{4}{|c|}{{ \bf{2$\ell$+0jets exclusive signal region definitions }}} \\ 
\hline
\mttwo\ [GeV] &  $m_{\ell\ell}$ [GeV]  &  SF bin & DF bin \\
\hline
\multirow{4}{*}{ 100--150} & 111--150 & SR2-SF-a & \multirow{4}{*}{SR2-DF-a} \\
& 150--200 & SR2-SF-b & \\
& 200--300 & SR2-SF-c & \\
& $>300$ & SR2-SF-d & \\
\hline
\multirow{4}{*}{ 150--200} & 111--150 & SR2-SF-e & \multirow{4}{*}{SR2-DF-b} \\
& 150--200 & SR2-SF-f & \\
& 200--300 & SR2-SF-g & \\
& $>300$ & SR2-SF-h & \\
\hline
\multirow{4}{*}{ 200--300} & 111--150 & SR2-SF-i & \multirow{4}{*}{SR2-DF-c} \\
& 150--200 & SR2-SF-j & \\
& 200--300 & SR2-SF-k & \\
& $>300$ & SR2-SF-l & \\
\hline
$>300$ & $>111$ & SR2-SF-m & SR2-DF-d \\
\hline
\multicolumn{4}{|c|}{{\bf{ 2$\ell$+0jets inclusive signal region definitions }}} \\ 
\hline
 $>$ 100 & $>$ 111 & SR2-SF-loose & -\\
 $>$ 130 & $>$ 300 &  SR2-SF-tight & -  \\
 \hline
 $>$ 100 & $>$ 111 &- & SR2-DF-100  \\
 $>$ 150 & $>$ 111 &- & SR2-DF-150\\
 $>$ 200 & $>$ 111 &- & SR2-DF-200\\
 $>$ 300 & $>$ 111 &- & SR2-DF-300 \\
\hline
\end{tabular}
\end{table}

\subsection{Signal regions for 2$\ell$+jets channel}

In the 2$\ell$+jets channel, two inclusive signal regions differing only in the \MET\ requirement, denoted SR2-int and SR2-high,
are used to target intermediate and large mass splittings between the $\chinoonepm /\ninotwo$ chargino/neutralino and the 
$\ninoone$ neutralino. In addition to the preselection used in the 2$\ell$+0jets channel, with the exception of the veto requirement on non-$b$-tagged jets, the sub-leading lepton is also required to have \pt$>25$ GeV and events must have at least two jets, with the leading two jets satisfying \pt$>30$ GeV. The $b$-jet veto is applied in the same way as in the 2$\ell$+0jets channel.
Several kinematic requirements are applied to select two leptons consistent with an on-shell $Z$ boson and two jets consistent with a $W$ boson. A tight requirement of $\mttwo>100$~GeV
is used to suppress the $t\bar{t}$ and $WW$ backgrounds
and $\MET>150~(250)$ GeV is required for SR2-int (SR2-high).

An additional region in the 2$\ell$+jets channel, denoted SR2-low, is optimized for the region of parameter space where the mass splitting
between the $\chinoonepm /\ninotwo$ and the $\ninoone$ is similar to the $Z$ boson mass and the signal becomes kinematically similar to the diboson ($VV$) 
backgrounds.
It is split into two orthogonal subregions for performing background estimation and validation, and these are merged when presenting the results in Section~\ref{sec:result}. SR2-low-2J requires exactly two jets, with $\pT>30$ GeV, that are both assumed to originate from the $W$ boson, while SR2-low-3J requires 3--5 signal jets (with the leading two jets satisfying \pt$>30$ GeV) and assumes the $\chinoonepm\ninotwo$ system recoils against initial-state-radiation (ISR) jet(s). In the latter case, the two jets originating from the $W$ boson are selected to be those closest in $\Delta \phi$ to the $Z(\rightarrow\ell\ell) + \MET$ system. This is different from SR2-int and SR2-high, where the two jets with the highest \pt in the event are used to define the $W$ boson candidate. The rest of the jets that are not associated with the $W$ boson are collectively defined as ISR jets.
All regions use variables, including angular distances and the $W$ and $Z$ boson transverse momenta, to select the signal topologies of interest.
The definitions of the signal regions in the 2$\ell$+jets channel are summarized in Table~\ref{tab:2L2J_SR_def}.

 \begin{table}[t]
   \centering
 \caption{
Signal region definitions used for the 2$\ell$+jets channel. Relevant kinematic variables are defined in the text. The symbols $W$ and $Z$ correspond to the reconstructed $W$ and $Z$ bosons in the final state. The $Z$ boson is always reconstructed from the two leptons, whereas the $W$ boson is reconstructed from the two jets leading in \pt for SR2-int, SR2-high and the 2-jets channel of SR2-low, whilst for the 3--5 jets channel of SR2-low it is reconstructed from the two jets which are closest in $\Delta \phi$ to the $Z$ ($\rightarrow\ell\ell$) + \MET system. The $\Delta R_{(jj)}$ and  $m_{jj}$ variables are calculated using the two jets assigned to the $W$ boson. ISR refers to the vectorial sum of the initial-state-radiation jets in the event (i.e. those not used in the reconstruction of the $W$ boson) and jet1 and jet3 refer to the leading and third leading jet respectively. The variable $n_{\text{non-} b \text{-tagged jets}}$ refers to the number of jets with $\pT>30$ GeV that do not satisfy the $b$-tagging criteria.
}
\label{tab:2L2J_SR_def}
   \begin{tabular}{|l|c|c|c|c| }
   \hline
     \multicolumn{5}{|c|}{\bf {2$\ell$+jets signal region definitions}} \\
     \hline
      &  SR2-int   &  SR2-high & SR2-low-2J &  SR2-low-3J \\
      \hline
         $n_{\text{non-} b \text{-tagged jets}}$     & \multicolumn{2}{c|}{$\geq 2$} & 2 & 3--5 \\[0.05cm]
        $m_{\ell\ell}$ [GeV]  & \multicolumn{2}{c|}{81--101} & 81--101  & 86--96 \\[0.05cm]
        $m_{jj}$ [GeV]      & \multicolumn{2}{c|}{70--100} &  70--90 & 70--90   \\[0.05cm]
        \MET [GeV]         &  $>150$ &$>250$ &  $>100$ & $>100$\\[0.05cm]

         $p^{Z}_{\rm T}$  [GeV]   & \multicolumn{2}{c|}{$>80$} & $>60$ & $>40$  \\[0.05cm]
         $p^{W}_{\rm T}$  [GeV]   & \multicolumn{2}{c|}{$>100$} &  &  \\[0.05cm]

        $m_{\rm T2}$ [GeV] &  \multicolumn{2}{c|}{$>100$} &  & \\[0.05cm]

         $\Delta R_{(jj)}$         & \multicolumn{2}{c|}{$<1.5$} &  &  $<2.2$ \\[0.05cm]
         $\Delta R_{(\ell\ell)}$   & \multicolumn{2}{c|}{$<1.8$} &  & \\[0.05cm]

        $\Delta\phi_{(\pTmiss ,Z)}$    & \multicolumn{2}{c|}{} & $<0.8$ &  \\[0.05cm]
        $\Delta\phi_{(\pTmiss ,W)}$    & \multicolumn{2}{c|}{0.5--3.0 } & $>1.5$  & $<2.2$ \\[0.05cm]
        
        $\MET/p^{Z}_{\rm T}$    & \multicolumn{2}{c|}{} &  $0.6--1.6$ &  \\[0.05cm]
        $\MET/p^{W}_{\rm T}$    & \multicolumn{2}{c|}{} &  $<0.8$ & \\[0.05cm]
        $\Delta\phi_{(\pTmiss ,\rm{ISR})} $ & \multicolumn{2}{c|}{} &   & $>2.4$\\[0.05cm]
        $\Delta\phi_{(\pTmiss ,\rm{jet1})}$ & \multicolumn{2}{c|}{} &   & $>2.6$ \\[0.05cm]
        
        $\MET/p^{\rm {ISR}}_{\rm T}$    & \multicolumn{2}{c|}{} &   & 0.4--0.8 \\[0.05cm]

        $|\eta(Z)|$    & \multicolumn{2}{c|}{} &   & $<1.6$ \\[0.05cm]
        $p^{\rm{jet3}}_{\rm T}$  [GeV]    & \multicolumn{2}{c|}{} &   & $>30$ \\[0.05cm]

   \hline
   \end{tabular}

\end{table}

\FloatBarrier

\subsection{Signal regions for 3$\ell$ channel}

The 3$\ell$ channel targets $\chinoonepm\ninotwo$ production and uses kinematic variables such as \MET and the transverse mass $\mT$, which were used in the Run 1 analysis~\cite{SUSY-2013-12}. 
Events are required to have exactly three signal leptons and no additional baseline leptons, as well as zero $b$-tagged jets with $\pT>20$ GeV. In addition, two of the leptons must form an SFOS pair (as expected in $\ninotwo\rightarrow \ell^{+}\ell^{-}\ninoone$ decays). 
To resolve ambiguities when multiple SFOS pairings are present, the transverse mass is calculated using the unpaired lepton and \pTmiss\ for each possible SFOS pairing, and the lepton that yields the minimum transverse mass is assigned to the $W$ boson. This transverse mass value is denoted by
$m_{\rm{T}}^{\rm{min}}$, and is 
used alongside \MET, jet multiplicity (in the gauge-boson-mediated scenario) and other relevant kinematic variables to define exclusive signal regions that have sensitivity to $\slepton$-mediated and gauge-boson-mediated decays. The definitions of these exclusive regions are provided in Table~\ref{tab:threelepSRs}. The bins denoted ``slep-a,b,c,d,e'' target $\slepton$-mediated decays and consequently have a veto on SFOS pairs with an invariant mass consistent with the $Z$ boson (this suppresses the $WZ$ background). The invariant mass of the SFOS pair, \mll, the magnitude of the missing transverse momentum, \MET, and the \pT\ value of the third leading lepton, $p_{\textrm{T}}^{\ell_{3}}$, are used to define the SR bins. 
Conversely, the bins denoted ``WZ-0Ja,b,c'' and ''WZ-1Ja,b,c'' target gauge-boson-mediated decays and thus require the SFOS pair to have an invariant mass consistent with an on-shell $Z$ boson. The 0-jet and $\geq$ 1-jet channels are considered separately and the regions are
binned in $\mT^{\rm{min}}$ and \MET.

\FloatBarrier
 \begin{table}[t]
   \centering
   \caption{Summary of the exclusive signal regions used in the 3$\ell$ channel. Relevant kinematic variables are defined in the text. The bins labelled ``slep'' target slepton-mediated decays whereas those labelled ``WZ'' target gauge-boson-mediated decays. The variable $n_{\text{non-} b \text{-tagged jets}}$ refers to the number of jets with $\pT>20$ GeV that do not satisfy the $b$-tagging criteria. Values of $p^{\ell_{3}}_{\rm T}$ refer to the \pt\ of the third leading lepton and $\pt^{\rm{jet1}}$ denotes the \pt\ of the leading jet.  \label{tab:threelepSRs}}
   \begin{tabular}{|c |c |c |c |c |c |c |c | }
   \hline
     \multicolumn{8}{|c|}{\bf{3$\ell$ exclusive signal region definitions}} \\
     \hline
        $m_{\rm{SFOS}}$  &  \MET  &   $p^{\ell_{3}}_{\rm T}$   &  $n_{\text{non-} b \text{-tagged jets}}$ & $\mT^{\rm{min}}$& $p^{\ell\ell\ell}_{\rm T}$  & $\pt^{\textup{jet1}}$  &  Bins \\
        $\rm{[GeV]}$        &   $\rm{[GeV]}$ &    $\rm{[GeV]}$  &                             &  $\rm{[GeV]}$        & $\rm{[GeV]}$                 &  $\rm{[GeV]}$ &  \\
     \hline
         \multirow{2}{*}{\textless 81.2}   & \multirow{2}{*}{$>130$} & 20--30   & \multirow{2}{*}{} & \multirow{2}{*}{$>110$} & \multirow{2}{*}{}&   \multirow{2}{*}{}  & SR3-slep-a    \\
                         &        & $>30$ &  &  & &  & SR3-slep-b \\
     \hline
       \multirow{3}{*}{\textgreater 101.2}      & \multirow{3}{*}{$>130$} &    20--50     &\multirow{3}{*}{} &\multirow{3}{*}{$>110$} & \multirow{3}{*}{}& \multirow{3}{*}{}  & SR3-slep-c  \\
                             &              &    50--80  &  & & &  & SR3-slep-d \\
                             &              &    $>80$  & & & & &  SR3-slep-e\\
     \hline

      \multirow{3}{*}{81.2--101.2}     & 60-120 &  & \multirow{3}{*}{0} & \multirow{3}{*}{$>110$}    & \multirow{3}{*}{} &  \multirow{3}{*}{}	   & SR3-WZ-0Ja \\
          & 120--170  &  &    &              & &    & SR3-WZ-0Jb    \\
      & $>170$    & 	 &           &  & &   & SR3-WZ-0Jc   \\
    \hline
       \multirow{3}{*}{81.2-101.2}   & 120--200                       &    & \multirow{3}{*}{$\geq$ 1} &   $>110$     & $<120$ & $>70$ & SR3-WZ-1Ja\\
                                     & \multirow{2}{*}{$>200$} &    &                          & 110--160            &  \multirow{2}{*}{}   &\multirow{2}{*}{} & SR3-WZ-1Jb\\
                                   &                               & $>35$ &                          & $>160$        &  &  & SR3-WZ-1Jc  \\
    
   \hline
   \end{tabular}

\end{table}

\section{Background estimation and validation}
\label{sec:backgrounds}
The SM backgrounds can be classified into irreducible backgrounds with prompt leptons and genuine \MET\ from neutrinos,
and reducible backgrounds that contain one or more ``fake'' or non-prompt (FNP) leptons or
where experimental effects (e.g., detector mismeasurement of jets or leptons or imperfect removal of object double-counting) lead to significant ``fake'' \MET.
A summary of the background estimation techniques used in each channel is provided in Table~\ref{tab:bgestimationmethods}.
In the 2$\ell$+0jets and 3$\ell$ channels only, the dominant backgrounds are estimated from MC simulation and normalized in dedicated control regions (CRs)
that are included, together with the SRs, in simultaneous likelihood fits to data, as described further in Section~\ref{sec:result}. In addition, all channels employ validation regions (VRs)
with kinematic requirements that are similar to the SRs but with smaller expected signal-to-background ratios, which are used to validate the background
estimation methodology. In the 2$\ell$+jets channel, the MC modelling of diboson processes is studied in dedicated VRs and found to accurately reproduce data.


\begin{table}[t]
\centering
\caption[]{Summary of the estimation methods used in each search channel. Backgrounds denoted CR have a dedicated control region that is included in a simultaneous likelihood fit to data to extract a data-driven normalization factor that is used to scale the MC prediction. The $\gamma$+jet template method is used in the 2$\ell$+jets channel to provide a data-driven estimate of the $Z$+jets background. Finally,
MC stands for pure Monte Carlo estimation.
}
\label{tab:bgestimationmethods}
\begin{tabular}{| c | c | c | c |}
\hline
 \multicolumn{4}{|c|}{ \bf{Background estimation summary}} \\
\hline
Channel & 2$\ell$+0jets & 2$\ell$+jets  & 3$\ell$\\
\hline
Fake/non-prompt leptons        &   \multicolumn{2}{c|}{Matrix method } & Fake-factor method \\
\hline
$t\bar{t} + Wt$     &  CR  & MC & Fake-factor method   \\ 
$VV$                &  CR & MC & CR (WZ-only)\\
$Z$+jets            &  MC & $\gamma$+jet template & Fake-factor method \\
\hline
Higgs/ $VVV$/ top+$V$               & \multicolumn{3}{c|}{MC}\\
\hline
\end{tabular}

\end{table}


For the  2$\ell$+0jets channel the dominant backgrounds are irreducible processes from SM diboson production ($WW$, $WZ$, and $ZZ$) and dileptonic $t\bar{t}$ and $Wt$ events.
MC simulation is used to predict kinematic distributions for these backgrounds, but the $t\bar{t}$ and diboson backgrounds are then normalized to data in dedicated control regions. For the diboson 
backgrounds, SF and DF events are treated separately and two control regions are defined. The first one (CR2-VV-SF) selects SFOS lepton pairs with an invariant mass consistent with the $Z$ boson mass and has a tight requirement of \mttwo\ $>130$ GeV to reduce the $Z$+jets contamination. This region is dominated by $ZZ$ events, with subdominant contributions from $WZ$ and $WW$ events. The DF diboson control region (CR2-VV-DF) selects events with a different flavour opposite sign pair and further requires $50<m_{\rm{T2}}<75$ GeV. This region is dominated by $WW$ events, with a subdominant contribution from $WZ$ events. The $t\bar{t}$ control region (CR2-Top) uses DF events with at least one $b$-tagged jet to obtain a high-purity sample of $t\bar{t}$ events. The control region definitions are summarized in Table~\ref{tab:CRdef_2LOS}.
The $Z$+jets and Higgs boson contributions are expected to be small in the 2$\ell$+0jets channel and are estimated directly from MC simulation.

\begin{figure}[!h]
  \centering
\begin{subfigure}[t]{0.48\textwidth}\includegraphics[width=\textwidth]{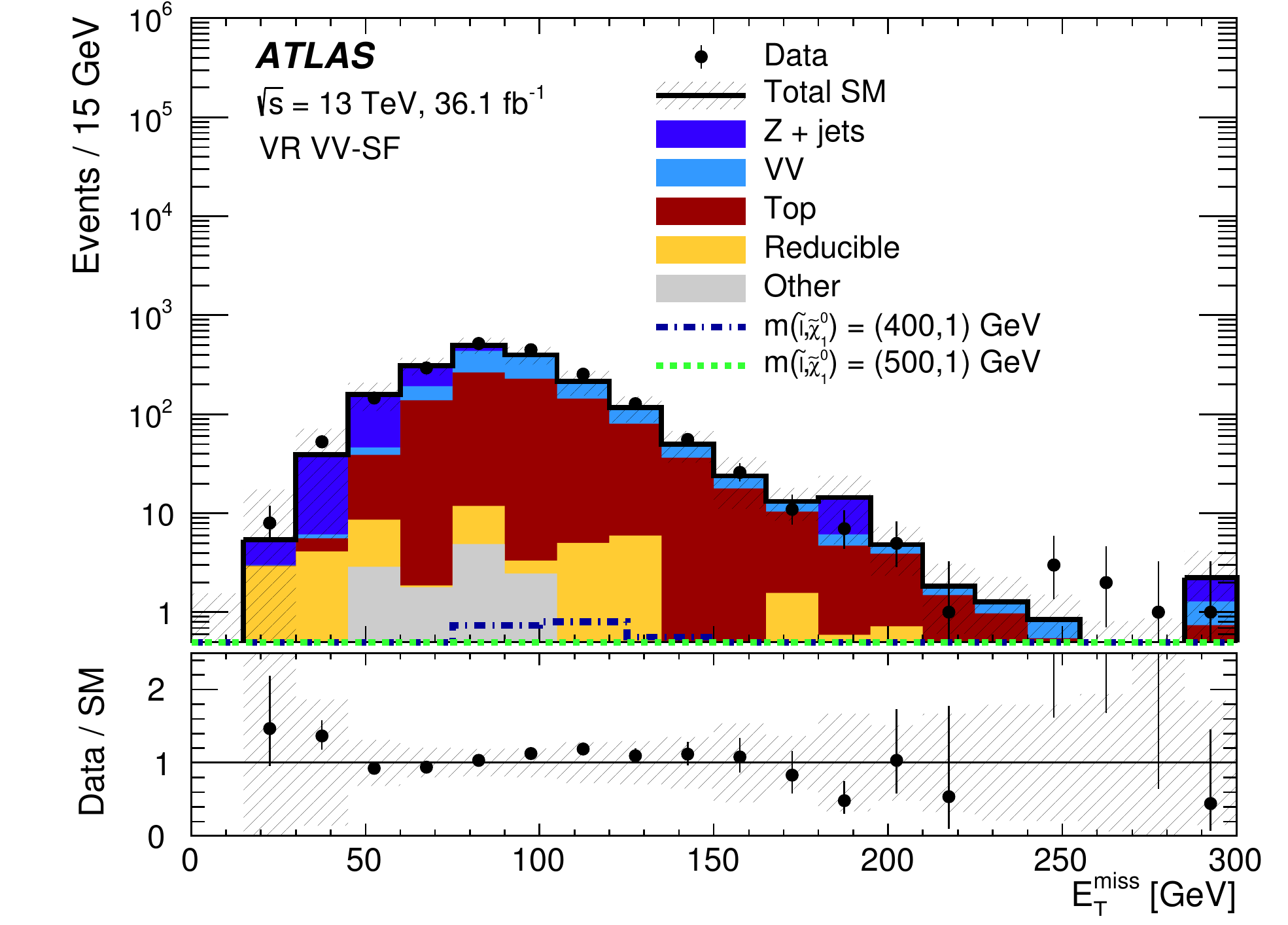}        
  \caption{\MET\ distribution in VR2-VV-SF}\label{figbody:metVRVVSF}\end{subfigure}
\begin{subfigure}[t]{0.48\textwidth}\includegraphics[width=\textwidth]{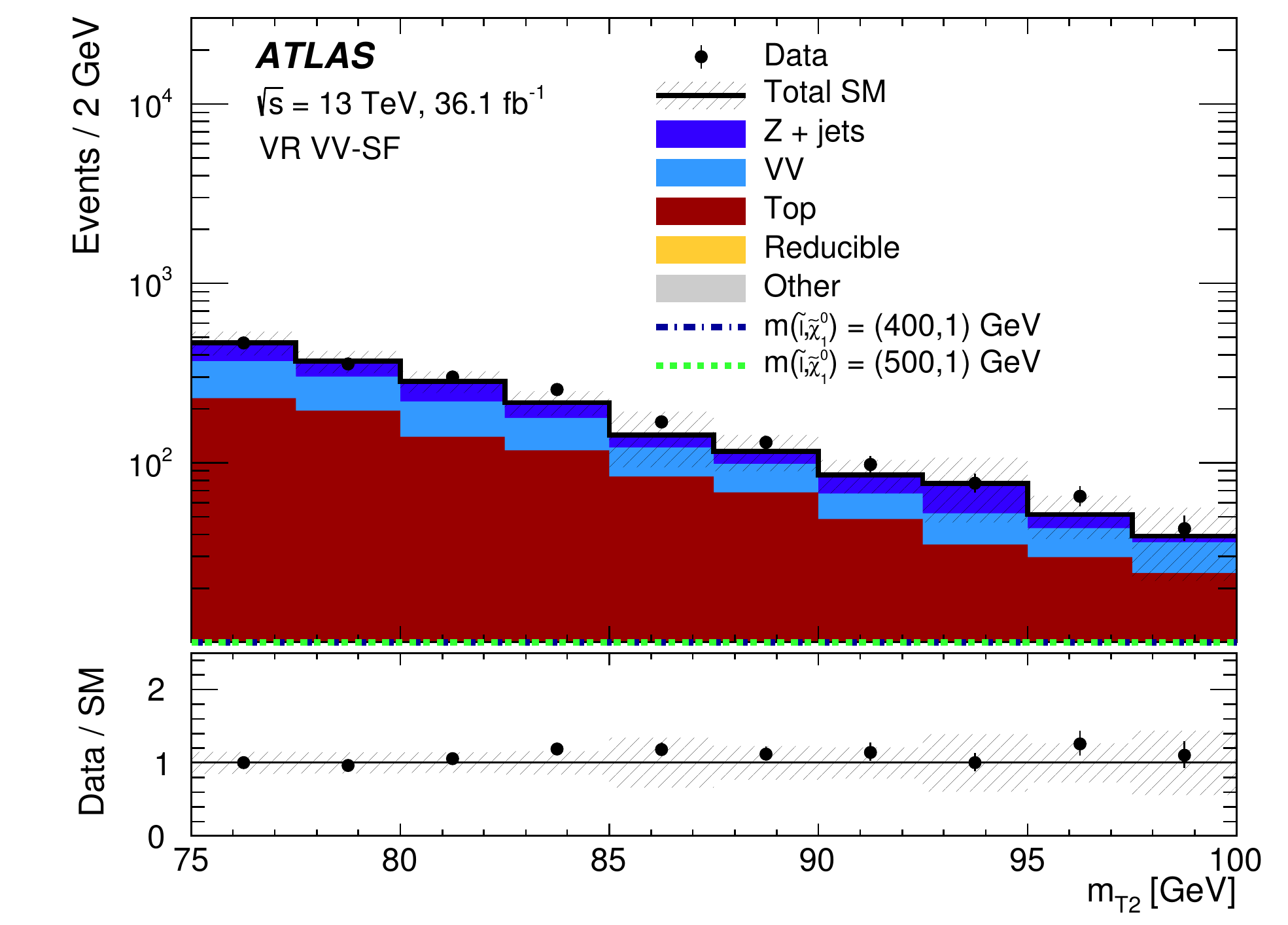}        
  \caption{\mttwo\ distribution in VR2-VV-SF}\label{figbody:mt2VRVVSF}\end{subfigure}
\begin{subfigure}[t]{0.48\textwidth}\includegraphics[width=\textwidth]{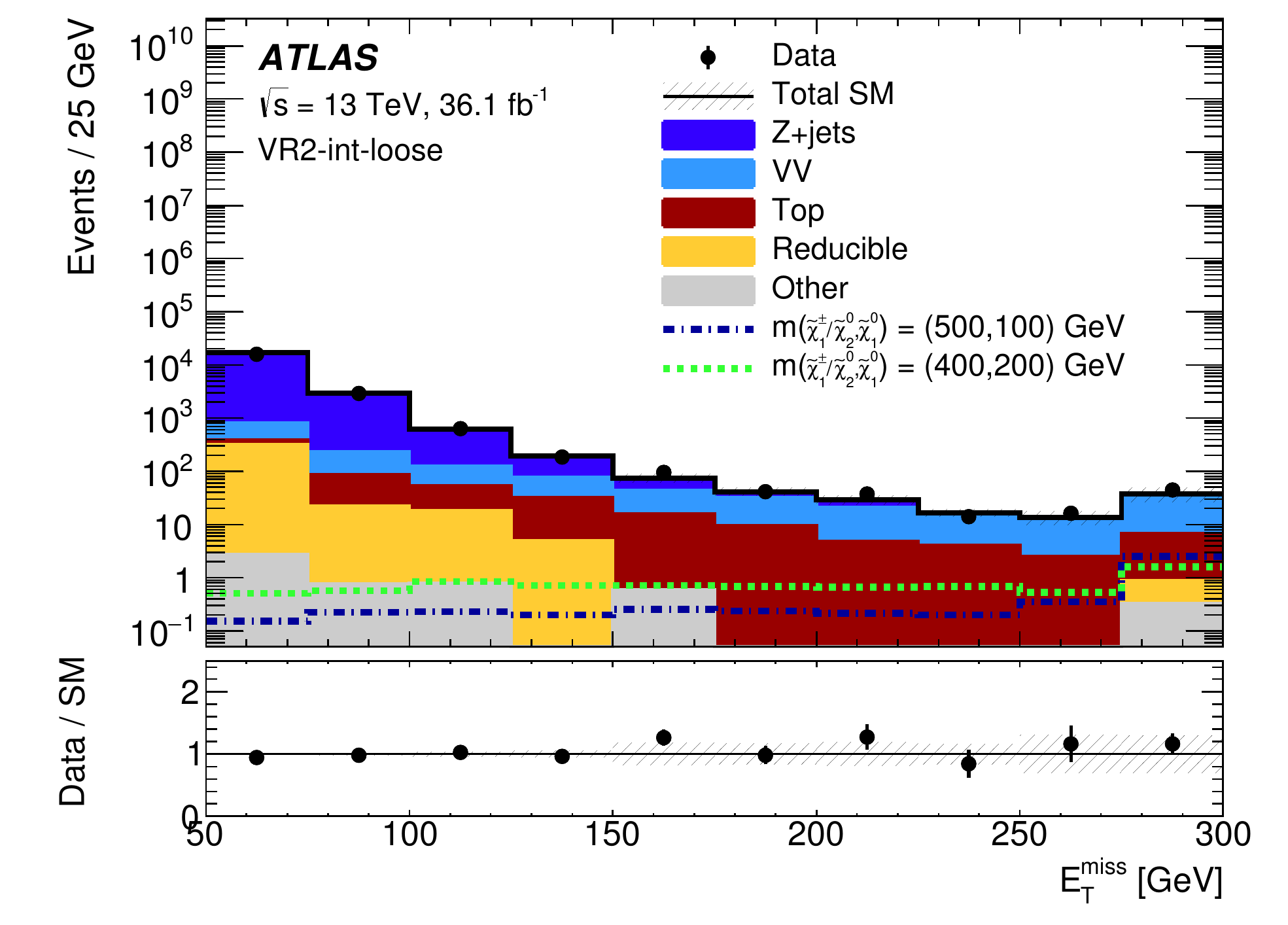}
  \caption{\MET\ distribution in VR2-int-loose}\label{figbody:metvr2int}\end{subfigure}
\begin{subfigure}[t]{0.48\textwidth}\includegraphics[width=\textwidth]{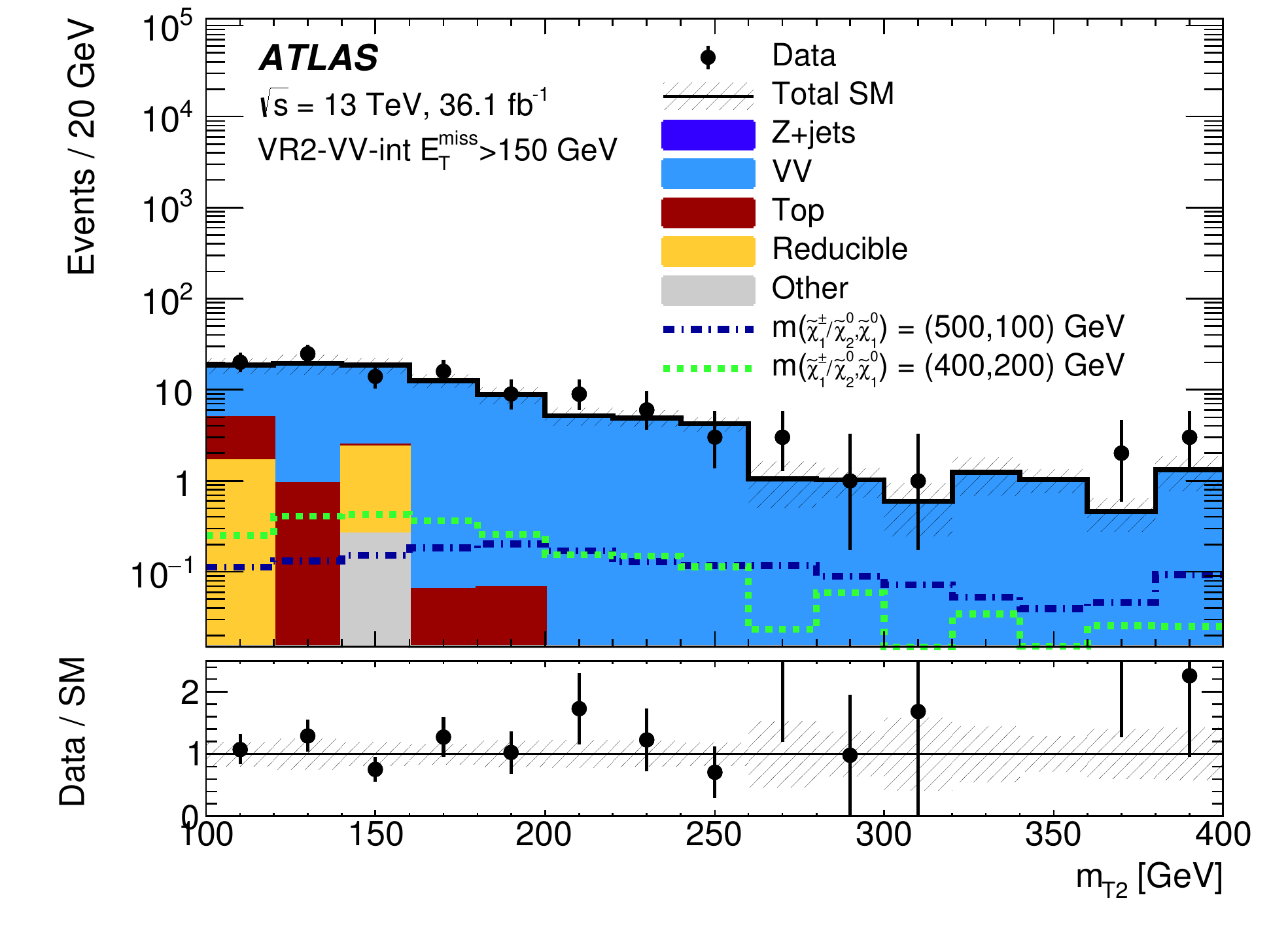}
  \caption{ \mttwo\ distribution in VR2-VV-int.}\label{figbody:mt2vr2vvint}\end{subfigure}
\begin{subfigure}[t]{0.48\textwidth}\includegraphics[width=\textwidth]{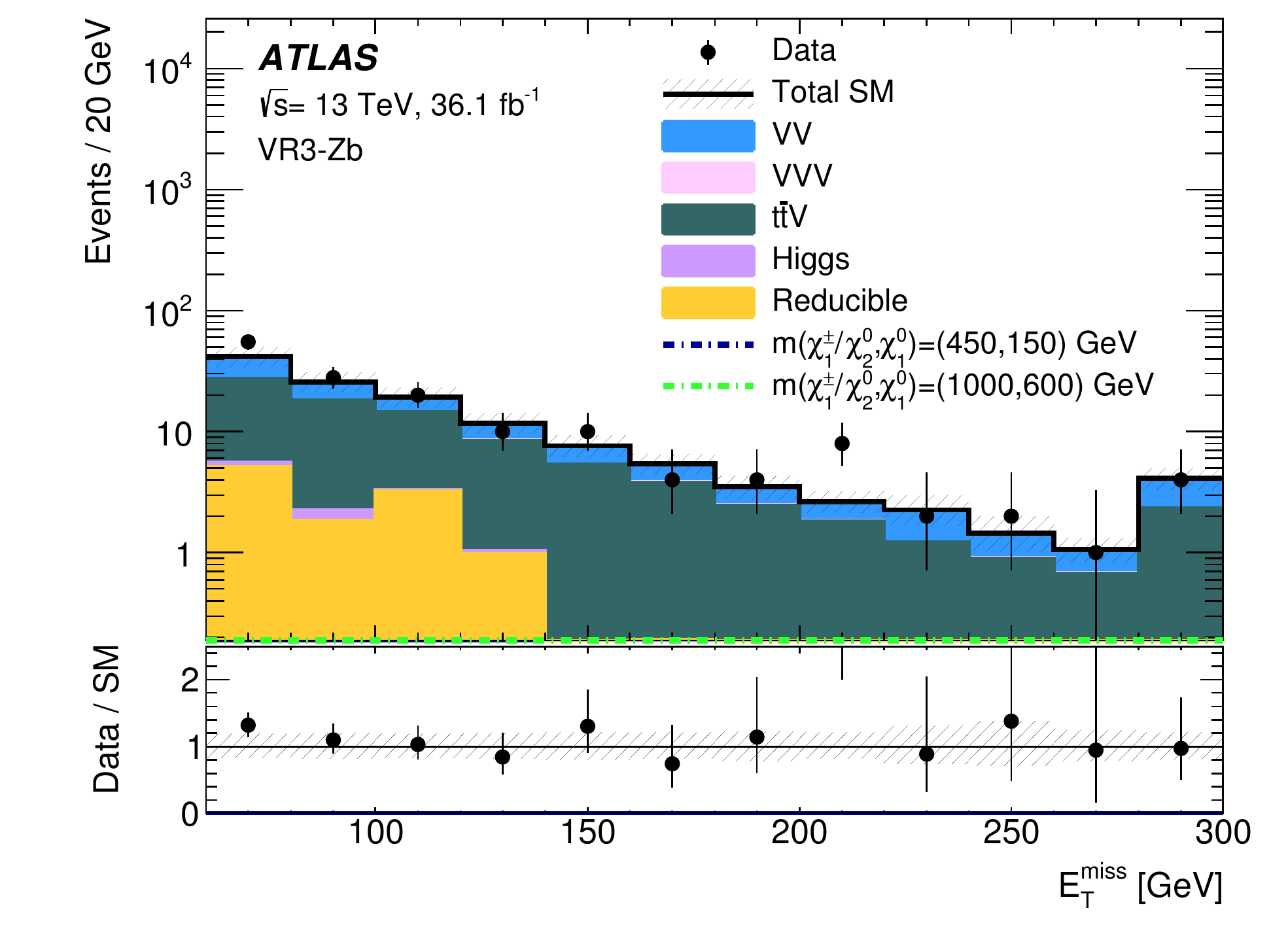}        
\caption{\MET\ distribution in VR3-Zb}\label{figbody:3lSlepBinBMET}\end{subfigure}
\begin{subfigure}[t]{0.48\textwidth}\includegraphics[width=\textwidth]{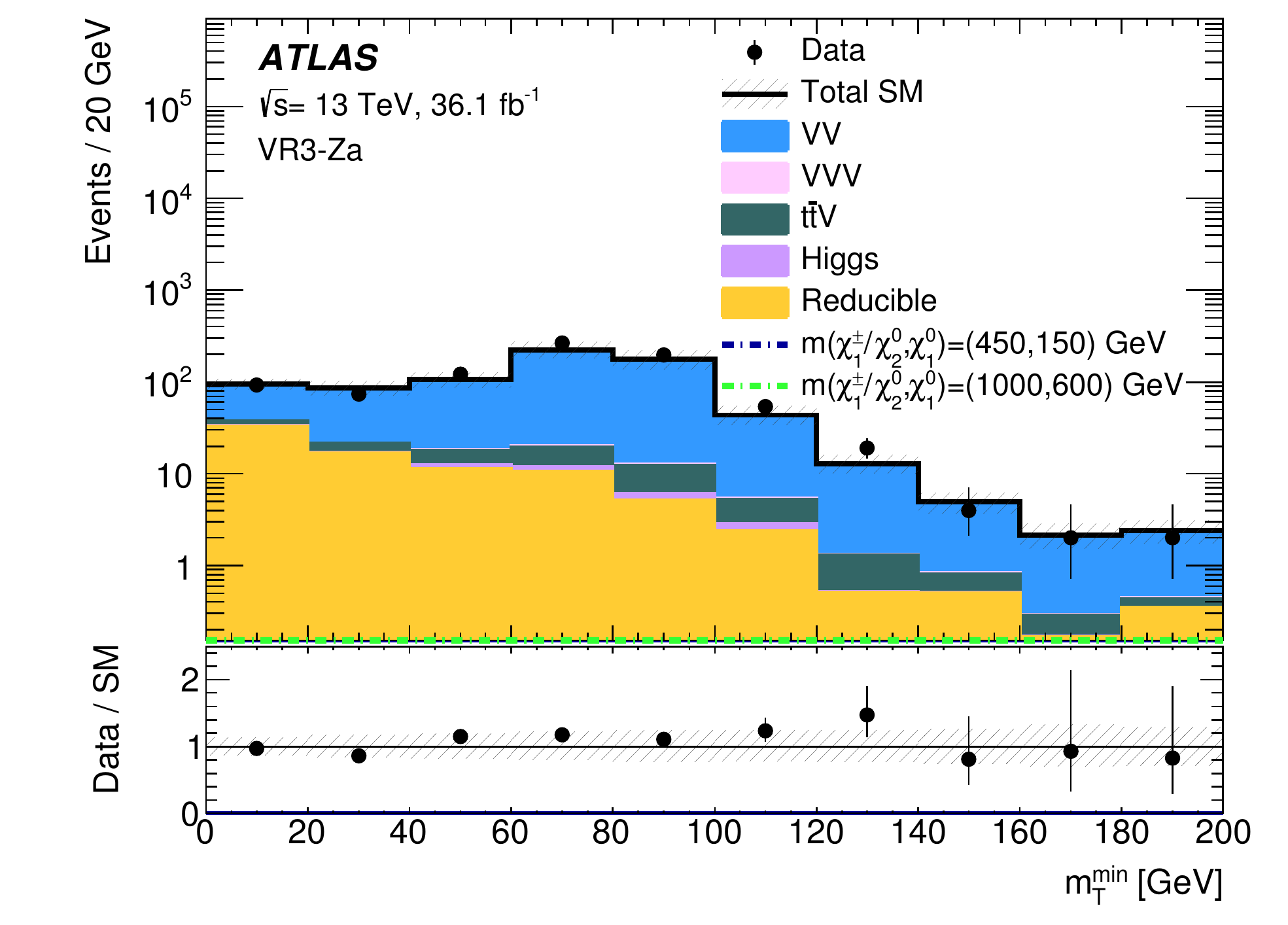} 
  \caption{$m_{T}^{\mathrm{min}}$ distribution in VR3-Za}\label{figbody:3lSlepBinAMET}\end{subfigure}
\caption{ Distributions of \MET, $m_{T}^{\mathrm{min}}$, and \mttwo\ for data and the estimated SM backgrounds in the (top) 2$\ell$+0jets channel,
(middle) 2$\ell$+jets channel, and (bottom) 3$\ell$ channel. Simulated signal models are overlaid for comparison. 
For the 2$\ell$+0jets (3$\ell$) channel, the normalization factors extracted from the corresponding CRs are used to rescale the $t\bar{t}$ and $VV$ ($WZ$) backgrounds. For the 2$\ell$+0jets channel the ``top'' background includes \ttbar and $Wt$, the ``other'' backgrounds include Higgs bosons, $t\bar{t}V$ and $VVV$ and the ``reducible'' category corresponds to the data-driven matrix method estimate.
For the 2$\ell$+jets channel, the ``top'' background includes \ttbar, $Wt$ and $t\bar{t}V$, the ``other'' backgrounds include Higgs bosons and $VVV$,
the ``reducible'' category corresponds to the data-driven matrix method estimate, and the $Z$+jets contribution is evaluated with 
the data-driven $\gamma$+jet template method.
For the 3$\ell$ channel, the ?reducible? category corresponds to the data-driven fake-factor estimate.
The uncertainty band includes all systematic and statistical sources and the final bin in each histogram also contains the events in the overflow bin.
}
\end{figure}

The three control regions are included in a simultaneous profile likelihood fit to the observed data which provides data-driven normalization factors for these backgrounds, as described in Section~\ref{sec:result}. The results are propagated 
to the signal regions, and to dedicated VRs that are defined in Table~\ref{tab:CRdef_2LOS}.
The normalization factors returned by the fit for the $t\bar{t}$, VV-DF and VV-SF backgrounds are $0.95\pm 0.03$, $1.06\pm 0.18$ and $0.96\pm 0.11$, respectively. Figures~\ref{figbody:metVRVVSF} and \ref{figbody:mt2VRVVSF} show the \MET and $m_{\mathrm{T2}}$ distributions, respectively, for data and
the estimated backgrounds in VR2-VV-SF with these normalization factors applied.

 \begin{table}[t]
 \begin{center}
 \caption{\label{tab:CRdef_2LOS} 
 Control region and validation region definitions for the 2$\ell$+0jets channel. The 
DF and SF labels refer to different-flavour or same-flavour lepton pair combinations, respectively.
The \pT\ thresholds placed on the requirements for $b$-tagged and non-$b$-tagged jets correspond to 20~GeV and 60~GeV, respectively. 
 }
 \begin{tabular}{|c|c|c|c|c|c|}
 \hline
 \multicolumn{6}{|c|}{ \bf{2$\ell$+0jets control and validation region definitions}} \\
 \hline
 Region                         & CR2-VV-SF & CR2-VV-DF &  CR2-Top &  VR2-VV-SF (DF)      & VR2-Top         \\
 \hline                                                   
 Lepton flavour                       & SF   & DF        & DF      &  SF (DF)             &  DF        \\
 $n_{\text{non-} b \text{-tagged jets}}$   & 0   & 0     &  0   &  0            &  0          \\
 $n_{b \text{-tagged jets}}$              & 0  &0 &   $\ge1$      &  $0$            & $\ge1$        \\
 $|m_{\ell\ell}-m_Z|$ [GeV]        & $<20$   & ---      & ---   & $>20$ (--)           &  ---    \\
 $m_{\mathrm{T2}}$ [GeV]        & $>130$   & 50--75   &  75--100     &  75--100    &  $>100$         \\
 \hline
 \end{tabular}
 \end{center}
 \end{table}

  In the 2$\ell$+jets channel, the largest background contribution is also from SM diboson production. In addition,
 $Z$+jets events can enter the SRs due to fake \MET\ from jet or lepton mismeasurements or genuine \MET\ from neutrinos in semileptonic decays of $b$- or $c$-hadrons. These effects are difficult to model in MC simulation, so instead $\Gjet$ events in data are used to extract the \MET\ shape in $\Zjet$ events, which have a similar topology and \MET\ resolution.
  Similar methods have been employed in searches for SUSY in events with two leptons, jets, and large \MET\ 
in ATLAS~\cite{SUSY-2016-05} 
and CMS~\cite{CMS-SUS-11-021,CMS-SUS-14-014}.
The \MET shape is extracted from a data control sample of $\Gjet$ events
using a set of single-photon triggers and weighting each event by the trigger prescale factor. 
Corrections to account for differences in the $\gamma$ and $Z$ boson $\pt$ distributions, as well as
different momentum resolutions for electrons, muons and photons, are applied.
Backgrounds of $W\gamma$ and $Z\gamma$ production, which contain a photon and genuine \MET\ from neutrinos, are subtracted using MC samples that are normalized to data in a $V\gamma$ control region containing a selected lepton and photon.
For each SR separately, the \MET\ shape is then normalized to data in a corresponding control region with $\MET<100$ GeV but all other requirements the same as in the SR.
To model quantities that depend on the individual lepton momenta, an \mll\ value is assigned to each \Gjet\ event by sampling from \mll\ distributions
(parameterized as functions of boson \pt\ and \METl, the component of \MET\ that is parallel to the boson's transverse momentum vector) extracted from a \Zjet\ MC sample. With this \mll\ value assigned to the photon, each \Gjet\ event is boosted to the rest frame of the hypothetical $Z$ boson and the photon is split into two pseudo-leptons, assuming isotropic decays in the rest frame.

  To validate the method, two sets of validation regions, ``tight'' and ``loose'', are defined for each SR. The definitions of these regions are provided in Table~\ref{tab:2L2J_VR_def}. 
The selections in the ``tight'' regions are identical to the SR selections with the exception of the dijet mass $m_{jj}$ requirement, which is replaced by the requirement ($m_{jj}<60$ GeV
  or $m_{jj}>100$ GeV) to suppress signal. These ``tight'' regions 
  are used to verify the expectation from the $\Gjet$ method that the residual $\Zjet$ background after applying the SR selections is very small.
The ``loose'' validation regions are instead defined by
  removing several other kinematic requirements used in the SR
  definition (\mttwo, all $\Delta\phi$ and $\Delta R$
  quantities, and the ratios of \MET\ to $W$ \pT, $Z$ \pT, and
   \pT of the system of ISR jets). These samples
  have enough $\Zjet$ events to perform comparisons of kinematic distributions, which validate
  the normalization and kinematic modelling of the $\Zjet$ background.
  The data distributions are consistent with the expected background in these validation regions, as shown in 
  Figure~\ref{figbody:metvr2int} for the \MET\ distribution in VR2-int-loose.

Once the signal region requirements are applied, the dominant background in the 2$\ell$+jets channel is the diboson background. This is taken from MC simulation, but the modelling is verified in two dedicated validation regions, one for signal regions with low mass-splitting (VR2-VV-low) and one for the intermediate and high-mass signal regions (VR2-VV-int).
Requiring high \MET\ and exactly one signal jet (compared to at least two jets in the SRs) suppresses the \ttbar\ background and enhances the purity of diboson events containing an ISR jet,
in which each boson decays leptonically.
Figure~\ref{figbody:mt2vr2vvint} shows the \mttwo\ distribution in VR2-VV-int for data and the expected backgrounds.

 \begin{table}[t]
   \centering
 \caption{Validation region definitions used for the 2$\ell$+jets channel. 
Symbols and abbreviations are analogous to those in Table~\ref{tab:2L2J_SR_def}.}
 \label{tab:2L2J_VR_def}
   \begin{tabular}{|c|c|c|c|c| }
   \hline
     \multicolumn{5}{|c|}{\bf{ 2$\ell$+jets validation region definitions}} \\
     \hline
      &  VR2-int(high) & VR2-low-2J(3J) &  VR2-VV-int & VR2-VV-low \\
      \hline

      \multicolumn{5}{|c|}{Loose selection} \\
      \hline
        $n_{\text{non-} b \text{-tagged jets}}$ &$\geq 2$ & 2 (3--5)& 1 & 1\\[0.05cm]
       
       \MET [GeV]         &  $>150$ ($>250$)  &   $>100$ &  $>150$ & $>150$\\[0.05cm]
            
         $m_{\ell\ell}$ [GeV]  & 81--101  &  81--101 (86--96) &  & 81--101\\[0.05cm]
         $m_{jj}$ [GeV]      & $\notin [60,100]$ & $\notin [60,100]$ & &  \\[0.05cm]

          $p^{Z}_{\rm T} $ [GeV]    & $>80$ & $>60$ ($>40$) &  & \\[0.05cm]
          $p^{W}_{\rm T} $ [GeV]    &  $>100$ &  &  &\\[0.05cm]

          $|\eta(Z)|$    &  &  $(<1.6 )$ &  &  \\
          $p^{\rm{jet}3}_{\rm T}$  [GeV]    &   & ($>30$) & &\\[0.05cm]

          $\Delta\phi_{(\pTmiss ,\rm{jet})}$    &  &   & $>0.4$ &  $>0.4$ \\[0.05cm]

          $m_{\rm T2}$ [GeV] &  &  &  $>100$ &\\[0.05cm]
          $\Delta R_{(\ell\ell)}$   &  &  & & $<0.2$\\[0.05cm]
          \hline
           \multicolumn{5}{|c|}{Tight selection} \\
           \hline                          
      
          $\Delta R_{(jj)}$         & $<1.5$ & ($<2.2$) & &\\[0.05cm]

          $\Delta\phi_{(\pTmiss ,W)}$    & 0.5--3.0 & $>1.5$ ($<2.2$) & &  \\[0.05cm]
          $\Delta\phi_{(\pTmiss ,Z)}$    &  & $ <0.8$ $(-)$ & & \\[0.05cm]

          $\MET/p^{W}_{\rm T}$    &  & $ <0.8$ $(-)$ & &  \\[0.05cm]
          $\MET/p^{Z}_{\rm T}$    &  &  0.6--1.6 $(-)$ & &  \\[0.05cm]
          $\MET/p^{\rm ISR}_{\rm T}$    &  &   (0.4--0.8) & &  \\[0.05cm]  

         $\Delta\phi_{(\pTmiss ,\rm{ISR})}$  &  & $(>2.4)$ & & \\[0.05cm]
         $\Delta\phi_{(\pTmiss ,\rm{jet1})}$  &  & $(>2.6)$ & & \\[0.05cm]

          $m_{\rm T2}$ [GeV]       & $>100$ &  &   &\\[0.05cm]
          $\Delta R_{(\ell\ell)}$   & $<1.8$ &  &   & \\[0.05cm]

   \hline
   \end{tabular}

\end{table}

For both the 2$\ell$+0jets and 2$\ell$+jets channels, reducible backgrounds with one or two FNP leptons arise from multijet, $W$+jets and single-top-quark production events.
For both analyses, the FNP lepton background is estimated from data using the matrix method (MM)~\cite{TOPQ-2010-01}. This method uses two types of lepton identification criteria: ``signal'', corresponding to leptons passing the full analysis selection, and ``baseline'',  corresponding to candidate electrons and muons as defined in Section~\ref{sec:objects}.
Probabilities for real leptons satisfying the baseline selection to also satisfy the signal selection are measured as a function of \pt\ and $\eta$ in dedicated regions enriched in $Z$ boson processes;
similar probabilities for FNP leptons are measured in events dominated by leptons from heavy flavour decays and photon conversions.
The method uses the number of observed events containing baseline--baseline, baseline--signal, signal--baseline and signal--signal lepton pairs in a given SR
to extract data-driven estimates for the FNP lepton background in the CRs, VRs, and SRs for each analysis. 

For the 3$\ell$ channel, the irreducible background is dominated by SM $WZ$ diboson processes. As in the 2$\ell$+0jets channel, the shape of this background is taken from MC simulation but normalized to data in a dedicated control region. The signal regions shown in Table~\ref{tab:threelepSRs} include a set of exclusive regions inclusive in jet multiplicity which target $\slepton$-mediated decays, and a set of exclusive regions separated into 0-jet and $\geq 1$ jet categories which target gauge-boson-mediated decays. To reflect this, three control regions are defined in order to extract the normalization of the $WZ$ background: an inclusive region (CR3-WZ-inc) and two exclusive control regions (CR3-WZ-0j and CR3-WZ-1j).  The results of the background estimations are validated in a set of dedicated validation regions.
This includes two validation regions that are binned in jet multiplicity (VR3-Za-0J and VR3-Za-1J),
and a set of inclusive validation regions (VR3-Za, VR3-Zb, VR3-offZa and VR3-offZb) targeting different regions of phase space considered in the analysis (i.e.\ within and outside the $Z$ boson mass window, high and low \MET, and vetoing events with a trilepton invariant mass within the $Z$ boson mass window). The definitions of the control and validation regions used in the 3$\ell$ analysis are shown in Table~\ref{tab:3lepCRVR}. The normalization factors extracted from the fit for inclusive $WZ$ events, $WZ$ events with zero jets, and $WZ$ events with at least one jet are $0.97\pm0.06$, $1.08\pm0.06$ and $0.94\pm0.07$, respectively. Other small 
background sources such as $VVV$, $tV$ and Higgs boson production processes contributing to the irreducible background are taken from MC simulation. 

\begin{table}[t]
\centering
   \caption{Control and validation region definitions used in the 3$\ell$ channel. 
     The $m_{\mathrm{SFOS}}$ quantity is the mass of the same-flavour opposite-sign lepton pair and $m_{\mathrm{\ell \ell \ell}}$ is the trilepton invariant mass.
   Other symbols and abbreviations are analogous to those in Table~\ref{tab:threelepSRs}.}  
  \label{tab:3lepCRVR}
  \setlength{\tabcolsep}{0.0pc}
  \begin{tabular*}{\textwidth}{@{\extracolsep{\fill}}|l  c c c c c  c c  |}
      \hline
      \multicolumn{8}{|c|}{\bf{ 3$\ell$ control and validation region definitions}} \\
      \hline
      ~        &  $p^{\ell_{3}}_{\rm T}$  & $m_{\mathrm{\ell \ell \ell}}$ & $m_{\mathrm{SFOS}}$  &            \MET  & $\mTmin$  & $n_{\text{non-} b \text{-tagged~jets}}$ & $n_{ b \text{-tagged jets}}$ \\
      ~        &  [GeV]  & [GeV] & [GeV]  &  [GeV]             & [GeV]  &   &  \\
            \hline
      CR3-WZ-inc & $>20$ &  -- & 81.2--101.2 &  $>120$ & $<110$ & --  & 0   \\
      CR3-WZ-0j  & $>20$ &  -- & 81.2--101.2 &   $>60$ & $<110$ & 0  &  0  \\
      CR3-WZ-1j  & $>20$ &  -- & 81.2--101.2 &   $>120$ & $<110$ & $>0$  & 0  \\

      \hline
      
      VR3-Za         &   $>30$   & \multirow{2}{*}{$\notin [81.2,101.2]$}          & 81.2--101.2        &  40--60    &--    & --       & --     \\
      VR3-Zb         &   $>30$   &       & 81.2--101.2        & $>$60      & --  &--     & $>0$         \\
   
      \hline
      VR3-offZa                &      $>30$     & \multirow{2}{*}{$\notin [81.2,101.2]$}      & \multirow{2}{*}{$\notin [81.2,101.2]$}          &   40--60  & --    &--    &--               \\
      VR3-offZb          & $>20$            &       &      & $>$ 40  & --  &--    & $>0$        \\

      \hline
      VR3-Za-0J                & \multirow{2}{*}{$>20$}    & \multirow{2}{*}{$\notin [81.2,101.2]$}           & \multirow{2}{*}{81.2--101.2}       & 40--60    & --    & 0   & 0          \\
      VR3-Za-1J               &          &       &    & 40--60    & --    & $>0$   & 0     
              \\
      \hline
    \end{tabular*}

\end{table}

In addition to processes contributing to the reducible backgrounds in the 2$\ell$ channels, the reducible backgrounds in the 3$\ell$ channel also include $Z+$jets, \ttbar, $WW$ and in general any physics process leading to less than three prompt and isolated leptons. The reducible backgrounds in the 3$\ell$ channel are estimated using a data-driven fake-factor (FF) method~\cite{STDM-2015-19}. This method uses two sets of lepton
identification criteria: the tight, or ``ID'', criteria corresponding to the signal lepton selection used in the analysis and the orthogonal loose, or ``anti-ID'', criteria which are designed to yield an enrichment in FNP leptons. In particular, for the anti-ID leptons the isolation and identification requirements applied to signal leptons are reversed. 
The $Z$+jets background events in the signal, control and validation regions are estimated using lepton \pT-dependent fake factors, defined as the ratio of the numbers of ID to anti-ID leptons in an FNP-dominated region. These fake factors are then applied to events passing selection requirements identical to those in the signal, control or validation region in question but where one of the ID leptons is replaced by an anti-ID lepton. The ``top-like'' contamination, which includes $t\bar{t}$, $Wt$, and $WW$, is subtracted from these anti-ID regions along with contributions from any remaining MC processes, to avoid double-counting. The top-like reducible background contributions are then estimated differently: data-to-MC scale factors derived with DF opposite-sign events are applied to simulated SF events. 
Figures~\ref{figbody:3lSlepBinBMET} and \ref{figbody:3lSlepBinAMET} show the \MET distribution in VR3-Zb and the $m_{T}^{\mathrm{min}}$ distribution in VR3-Za, respectively.

\clearpage

\section{Systematic uncertainties}
\label{sec:systematics}
Several sources of experimental and theoretical systematic uncertainty are considered in the SM background estimates and signal
predictions. These uncertainties are included in the profile likelihood fit described
in Section~\ref{sec:result}. The primary sources of systematic uncertainty are related to the jet energy scale (JES) and resolution (JER), theory uncertainties in the MC modelling, the reweighting procedure applied to simulation to match the distribution of the number of reconstructed vertices observed in data, the systematic uncertainty considered in the non-prompt background estimation and the theoretical cross-section uncertainties. The statistical uncertainty of the simulated event samples is taken into account as well. The effects of these uncertainties were evaluated for all signal samples and background processes. In the 2$\ell$+0jets and 3$\ell$ channels the normalizations of the MC predictions for the dominant background processes are extracted in dedicated control regions and the systematic uncertainties thus only affect the extrapolation to the signal regions in these cases. 

The JES and JER uncertainties are derived as a function of jet \pT and $\eta$, as well as of the pile-up conditions and the jet flavour composition of the selected jet sample. They are determined using a combination of data and simulation, through measurements of the jet response balance in dijet, $Z$+jets and $\gamma+$jets events \cite{ATL-PHYS-PUB-2015-015,Aaboud:2017jcu}.

The systematic uncertainties related to the \MET modelling in the simulation are estimated by propagating the uncertainties in the energy or momentum scale of each of the physics objects, as well as the uncertainties in the soft term's resolution and scale~\cite{ATL-PHYS-PUB-2015-023}.

The remaining detector-related systematic uncertainties, such as those in the lepton reconstruction efficiency, energy scale and energy resolution, in the $b$-tagging efficiency and in the modelling of the trigger~\cite{ATL-PHYS-PUB-2015-041,PERF-2015-10}, are included but were found to be negligible in all channels.

The uncertainties coming from the modelling of diboson events in MC simulation are estimated by varying the renormalization, factorization and merging scales used to generate the samples, and the PDFs. 
In the 2$\ell$+0jets channel the impact of these uncertainties in the modelling of $Z$+jets events is also considered, as well as 
uncertainties 
in the modelling of $t\bar{t}$ events due to parton shower simulation 
(by comparing samples generated with \POWHEG\ +\ \PYTHIA to \POWHEG\ +\ \Herwigpp~\cite{Bahr:2008pv}), ISR/FSR modelling 
(by comparing the predictions from an event sample generated by \POWHEG\ +\ \PYTHIA with those from two samples where the radiation settings are varied), and the PDF set. 

In the 2$\ell$+jets channel, uncertainties in the data-driven $\Zjet$ estimate are calculated following the methodology used in Ref.~\cite{SUSY-2016-05}. An additional uncertainty is based on the difference between the expected background yield from the nominal method and a second method implemented as a cross-check, which extracts the dijet mass shape from data validation regions, normalizes the shape to the sideband regions of the SRs, and extrapolates the background into the $W$ mass region.

For the matrix-method and fake-factor estimates of the FNP background, systematic uncertainties are assigned to account for 
differences in FNP lepton composition between the SR and the CR used to derive the fake rates and fake factors.
An additional uncertainty is assigned to the MC subtraction of prompt leptons from this CR.

The exclusive SRs in the 2$\ell$+0jets and 3$\ell$ channels are dominated by statistical uncertainties in the background estimates (which range from 10\% to 70\% in the higher mass regions in the 2$\ell$+0jets channel and from 5\% to 30\% in the 3$\ell$ channel). The largest systematic uncertainties are those related to diboson modelling, the JES and JER uncertainties and those associated with the \MET\ modelling.
In the 2$\ell$+jets channel the dominant uncertainties are those associated with the data-driven estimate of the $\Zjet$ background, which range from approximately 45\% to 75\%.

\FloatBarrier

\section{Results}
\label{sec:result}

The HistFitter framework~\cite{HistFitter} is used for the statistical interpretation of the results, with the CRs (for the 2$\ell$+0jets and 3$\ell$ channels) and SRs both participating in a simultaneous likelihood fit. The likelihood is built as the product of a Poisson probability density function describing the observed number of events in each CR/SR and Gaussian distributions that constrain the nuisance parameters associated with the systematic uncertainties and whose widths correspond to the sizes of these uncertainties; Poisson distributions are used instead for MC statistical uncertainties. Correlations of a given nuisance parameter among the different background sources and the signal are taken into account when relevant. 

\begin{figure}[htb!]
\centering
\includegraphics[width=1.0\textwidth]{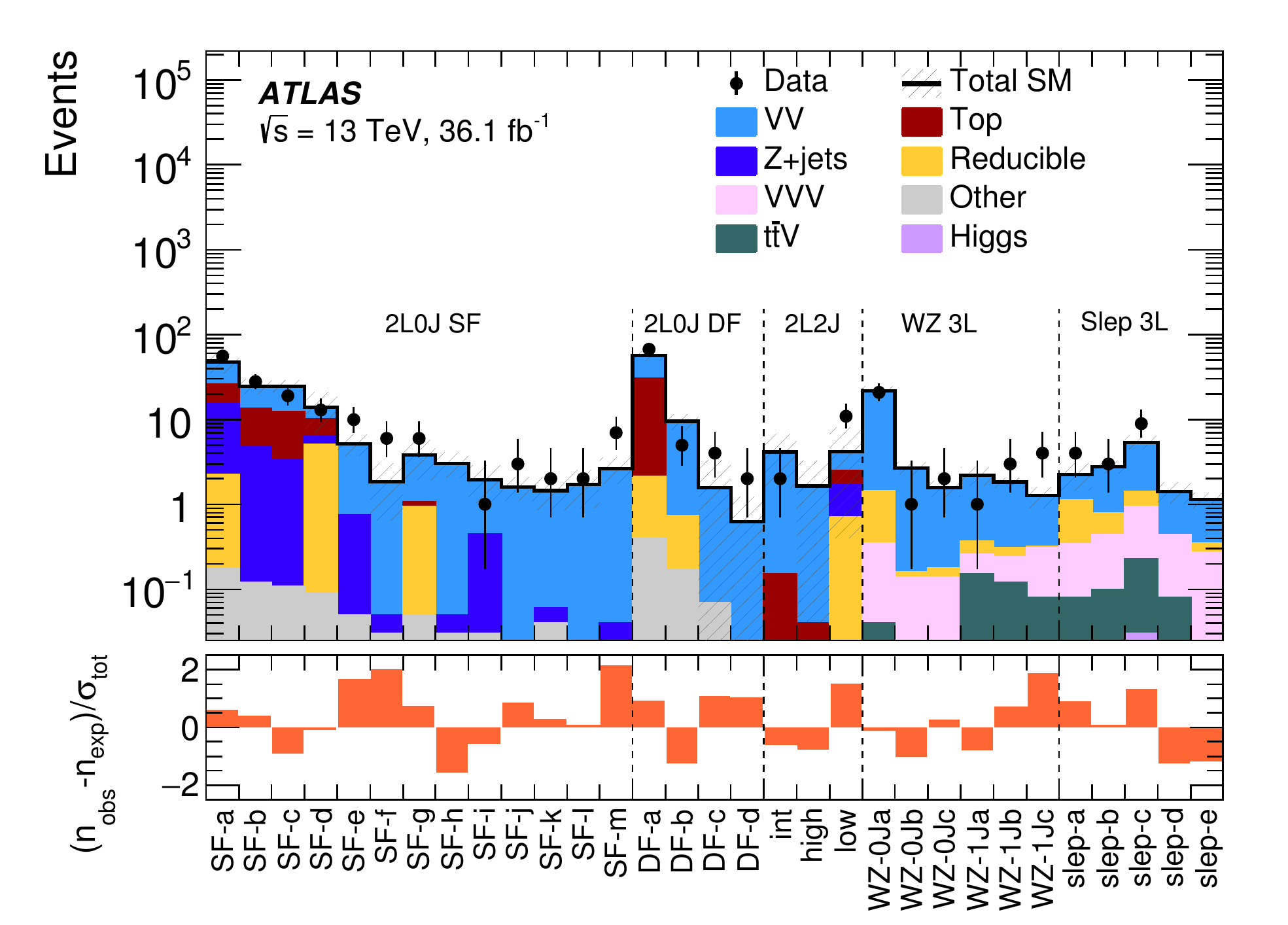}

\caption{The observed and expected SM background yields in the signal regions considered in the 2$\ell$+0jets, 2$\ell$+jets and 3$\ell$ channels. The statistical uncertainties in the background prediction are included in the uncertainty band, together with the experimental and theoretical uncertainties. The bottom plot shows the difference in standard deviations between the observed and expected yields.
Here $n_{\mathrm{obs}}$ and $n_{\mathrm{exp}}$ are the observed data and expected background yields, respectively,
$\sigma_{\mathrm{tot}}=\sqrt{n_{\mathrm{bkg}}+\sigma^{2}_{\mathrm{bkg}}}$, and $\sigma_{\mathrm{exp}}$ is the total background uncertainty. }
\label{fig:summaryPlot}
\end{figure}

In the 2$\ell$+0jets and 3$\ell$ channels, a background-only fit which uses data in the CRs is performed to constrain the nuisance parameters of the likelihood function (these include the normalization factors for dominant backgrounds and the parameters associated with the systematic uncertainties). In all channels the background estimates are also used to evaluate how well the expected and observed numbers of events agree in the validation regions, and good agreement is found. 
In the 2$\ell$+0jets, 2$\ell$+jets, and 3$\ell$ channels, the number of considered VRs is 3, 8, and 6, respectively, 
and the most significant deviations observed are 0.4$\sigma$, 1.4$\sigma$, and 0.8$\sigma$, respectively.
The precision of the expected background yields in the VRs is significantly better than in the corresponding SRs
and the dominant sources of systematic uncertainty in the VRs and corresponding SRs are similar.
For the 2$\ell$+0jets channel, the results for the exclusive signal regions are shown in Tables~\ref{tab:SR0jetResultsA}, \ref{tab:SR0jetResultsB}	and~\ref{tab:SR0jetResultsC} for SR2-SF-a to SR2-SF-g, SR2-SF-h to SR2-SF-m and SR2-DF-a to SR2-DF-d, respectively. The results for the 2$\ell$+0jets inclusive signal regions are shown in Table~\ref{tab:SR0jetInc}, while Table~\ref{tab:SR2LjetResults} summarizes the expected SM background and observed events in the 2$\ell$+jets SRs. For the 3$\ell$ channel, the results are shown in Table~\ref{tab:SR3LResultsWZ} for SR3-WZ-0Ja to SR3-WZ-0Jc and SR3-WZ-1Ja to SR3-WZ-1Jc (which target gauge-boson-mediated decays) and Table~\ref{tab:SR3LResultsSlep} for SR3-slep-a to SR3-slep-e. A summary of the observed and expected yields in all of the signal regions considered in this paper is provided in Figure~\ref{fig:summaryPlot}. No significant excess above the SM expectation is observed in any SR.

\begin{table}[t]
\centering
\caption{Background-only fit results for SR2-SF-a to  SR2-SF-g in the 2$\ell$+0jets channel. All systematic and statistical uncertainties are included in the fit. The ``other'' backgrounds include all processes producing a Higgs boson, $VVV$ or $t\bar{t}V$. A ``--'' symbol indicates that the background contribution is negligible.}
\label{tab:SR0jetResultsA}
\setlength{\tabcolsep}{0.0pc}
{\small
\begin{tabular*}{\textwidth}{@{\extracolsep{\fill}}lccccccc}
\noalign{\smallskip}\hline\noalign{\smallskip}
SR2-           & SF-a            & SF-b            & SF-c            & SF-d            & SF-e            & SF-f            & SF-g              \\[-0.05cm]
\noalign{\smallskip}\hline\noalign{\smallskip}
Observed           & $56$              & $28$              & $19$              & $13$              & $10$              & $6$              & $6$                    \\
\noalign{\smallskip}\hline\noalign{\smallskip}
Total SM          &$ 47  \pm  12 $ & $ 25  \pm  5\;\,\,$ & $ 25  \pm  4\;\,\, $ & $ 14  \pm  7\;\,\, $ & $ 5.2  \pm  1.4 $ & $ 1.9  \pm  1.2 $ & $ 3.8  \pm  1.9 $ \\
\noalign{\smallskip}\hline\noalign{\smallskip}
         $t\bar{t}$          & $ 10  \pm  4\;\, $ & $ 7.4  \pm  3.5 $ & $ 7.3  \pm  3.0 $ & $ 2.7  \pm  1.7 $       & --          & --          & $0.11_{-0.11}^{+0.21}\,\,$              \\
         $Wt$          & $ \,1.0  \pm  1.0 $ & $ 1.3  \pm  0.7 $ & $ 1.6  \pm  0.6 $     &  $ 1.1  \pm  1.1 $           & --          & --          & --              \\
         $VV$      &   $ 21  \pm  4\;\, $ & $ 11.3  \pm  2.9\;\, $ & $ 12.6  \pm  2.4\;\, $ & $ 3.9  \pm  2.4 $ & $ 4.4  \pm  1.3 $ & $ 1.8  \pm  1.2 $ & $ 2.8  \pm  1.6 $            \\
         FNP          & $2.1_{-2.1}^{+2.9}\,$  & $0.0^{+0.4}_{-0.0}$  & $0.0^{+0.5}_{-0.0}$      & $5 \pm 4$          & $0.0^{+0.1}_{-0.0}$  & $0.00^{+0.01}_{-0.00}$          & $0.9 \pm 0.4$              \\
         $Z$+jets          & $ 13  \pm  9\;\, $ & $ 4.7  \pm  2.6 $ & $ 3.3  \pm  3.2 $    & $1.2_{-1.2}^{+1.7}\,$          & $ 0.7  \pm  0.6 $           & $0.02_{-0.02}^{+0.21}\,\,$          & --              \\
         Other          & $ \,0.18  \pm  0.08 $ & $ 0.12  \pm  0.05 $ & $ 0.11  \pm  0.04 $ & $ 0.09  \pm  0.05 $ & $ 0.05  \pm  0.03 $ & $ 0.03  \pm  0.01 $ & $ 0.05  \pm  0.02 $             \\

\noalign{\smallskip}\hline\noalign{\smallskip}
\end{tabular*}
}

\end{table}

\begin{table}[t]
\centering
\caption{Background-only fit results for SR2-SF-h to  SR2-SF-m in the 2$\ell$+0jets channel. All systematic and statistical uncertainties are included in the fit. The ``other'' backgrounds include all processes producing a Higgs boson, $VVV$ and $t\bar{t}V$.  A ``--'' symbol indicates that the background contribution is negligible.}
\label{tab:SR0jetResultsB}
\setlength{\tabcolsep}{0.0pc}
{\small
\begin{tabular*}{\textwidth}{@{\extracolsep{\fill}}lcccccc}
\noalign{\smallskip}\hline\noalign{\smallskip}
SR2-           & SF-h            & SF-i            & SF-j            & SF-k            & SF-l            & SF-m              \\[-0.05cm]
\noalign{\smallskip}\hline\noalign{\smallskip}
Observed           & $0$              & $1$              & $3$              & $2$              & $2$              & $7$                    \\
\noalign{\smallskip}\hline\noalign{\smallskip}
Total SM          & $ 3.1  \pm  1.0 $ & $ 1.9  \pm  0.9 $ & $ 1.6  \pm  0.5 $ & $ 1.5  \pm  0.6 $ & $ 1.8  \pm  0.8 $ & $ 2.6  \pm  0.9 $             \\
\noalign{\smallskip}\hline\noalign{\smallskip}
        $t\bar{t}$          & --          & --          & --          & --          & --          & --              \\
        $Wt$          & --          & --          & --          & --          & --          & --              \\
        $VV$          &$ 3.0  \pm  1.0 $ & $ 1.5  \pm  0.8 $ & $ 1.6  \pm  0.5 $ & $ 1.4  \pm  0.6 $ & $ 1.7  \pm  0.8 $ & $ 2.6  \pm  0.9 $             \\
        FNP          & $0.00^{+0.02}_{-0.00}$  &  $0.0^{+0.1}_{-0.0}$  &  $0.00^{+0.01}_{-0.00}$  &  $0.00^{+0.01}_{-0.00}$ &  $0.00^{+0.02}_{-0.00}$ &  $0.00^{+0.01}_{-0.00}$              \\
        $Z$+jets          & $0.02_{-0.02}^{+0.11}\,\,$          & $0.42 \pm 0.20$          & --          & $0.02_{-0.02}^{+0.20}\,\,$          & --          & $0.02_{-0.02}^{+0.06}\,\,$              \\
        Other          & $0.03 \pm 0.01$          & $0.03 \pm 0.02$           & --        & $0.04 \pm 0.02$          & $0.02 \pm 0.01$          & $0.02 \pm 0.02$              \\
\noalign{\smallskip}\hline\noalign{\smallskip}
\end{tabular*}
}

\end{table}

\begin{table}[t]
\centering
\caption{Background-only fit results for SR2-DF-a to  SR2-DF-d in the 2$\ell$+0jets channel. All systematic and statistical uncertainties are included in the fit.  The ``other'' backgrounds include all processes producing a Higgs boson, $VVV$ or $t\bar{t}V$.  A ``--'' symbol indicates that the background contribution is negligible.}
\label{tab:SR0jetResultsC}
\setlength{\tabcolsep}{0.0pc}
{\small

 \begin{tabular*}{\textwidth}{@{\extracolsep{\fill}}lcccc}
 \noalign{\smallskip}\hline\noalign{\smallskip}
 SR2-           & DF-a            & DF-b            & DF-c            & DF-d              \\[-0.05cm]
 \noalign{\smallskip}\hline\noalign{\smallskip}
 Observed           & $67$              & $5$              & $4$              & $2$                    \\
 \noalign{\smallskip}\hline\noalign{\smallskip}
Total SM         & $ 57  \pm  7\;\,\, $ & $ 9.6  \pm  1.9 $         & $1.5_{-1.5}^{+1.7}\,\,$          & $0.6 \pm 0.6$              \\
 \noalign{\smallskip}\hline\noalign{\smallskip}
         $t\bar{t}$          & $ 24  \pm  8\;\, $          & --          & --          & --              \\
         $Wt$          & $4.5 \pm 1.0$          & --          & --          & --              \\
         $VV$          & $ 26  \pm  6\;\, $ & $ 8.8  \pm  1.8 $     & $1.5_{-1.5}^{+1.7}\,\,$          & $0.6 \pm 0.6$              \\
         FNP          &$ 1.75  \pm  0.18 $ & $ 0.57  \pm  0.23 $ &  $0.00^{+0.01}_{-0.00}$  &  $0.00^{+0.01}_{-0.00}$              \\
         $Z$+jets          & --          & --          & --          & --              \\
         Other          &$ 0.40  \pm  0.09 $ & $ 0.17  \pm  0.07 $ & $ 0.07  \pm  0.07 $ & $ 0.02  \pm  0.02 $         \\
\noalign{\smallskip}\hline\noalign{\smallskip}
\end{tabular*}
}

\end{table}

\begin{table}
\centering
\caption{Background-only fit results for the inclusive signal regions in the 2$\ell$+0jets channel. All systematic and statistical uncertainties are included in the fit.  The ``other'' backgrounds include all processes producing a Higgs boson, $VVV$ and $t\bar{t}V$.  A ``--'' symbol indicates that the background contribution is negligible. }
\label{tab:SR0jetInc}
\setlength{\tabcolsep}{0.0pc}
{\small
\begin{tabular*}{\textwidth}{@{\extracolsep{\fill}}lcccccc}
\noalign{\smallskip}\hline\noalign{\smallskip}
SR2-           & SF-loose            & SF-tight             & DF-100            & DF-150            & DF-200           & DF-300                     \\[-0.05cm]
\noalign{\smallskip}\hline\noalign{\smallskip}
Observed           & $153$              & $9$              & $78$              & $11$              & $6$              & $2$                    \\
\noalign{\smallskip}\hline\noalign{\smallskip}
Total SM          &$ 133  \pm  22\;\,\, $ & $ 9.8  \pm  2.9 $ & $ 68  \pm  7\;\,\, $ & $ 11.5  \pm  3.1\;\,\, $ & $ 2.1  \pm  1.9 $ & $ 0.6  \pm  0.6 $        \\
\noalign{\smallskip}\hline\noalign{\smallskip}
        $t\bar{t}$          & $27 \pm 11$          & --          & $24 \pm 8\;\,$          & --           &--          & --           \\
        $Wt$          & $5.0 \pm 2.2$          &--         & $4.5 \pm 1.0$          & --        &--          & --            \\
        $VV$          &$ 70  \pm  11 $ & $ 9.6  \pm  3.0 $ & $ 37  \pm  8\;\, $ & $ 10.8  \pm  3.0\;\, $ & $ 2.0  \pm  1.9 $          & $0.6 \pm 0.6$              \\
        FNP          & $ 6  \pm  4 $ & $ 0.0  \pm  0.0 $ & $ 2.17  \pm  0.29 $ & $ 0.42  \pm  0.23 $       &  $0.00^{+0.01}_{-0.00}$           &  $0.00^{+0.01}_{-0.00}$             \\
        $Z$+jets          & $23 \pm 14$          & $0.09_{-0.09}^{+0.34}\,\,$          &--           & --           & --          & --               \\
        Other          & $ 0.79  \pm  0.23 $ & $ 0.09  \pm  0.01 $ & $ 0.67  \pm  0.16 $ & $ 0.26  \pm  0.08 $ & $ 0.09  \pm  0.07 $ & $ 0.02  \pm  0.02 $           \\
\noalign{\smallskip}\hline\noalign{\smallskip}
\end{tabular*}
}

\end{table}

\begin{table}[t]
\centering
\caption{ SM background results in the 2$\ell$+jets SRs. All systematic and statistical uncertainties are included. The ``top'' background includes
all processes producing one or more top quarks and the
``other'' backgrounds include all processes producing a Higgs boson or $VVV$.  A ``--'' symbol indicates that the background contribution is negligible. }
\label{tab:SR2LjetResults}
\setlength{\tabcolsep}{0.0pc}
{\small

\begin{tabular*}{\textwidth}{@{\extracolsep{\fill}}lccc}
\noalign{\smallskip}\hline\noalign{\smallskip}
		SR2-	& int		& high		& low (combined) \\[-0.05cm]
\noalign{\smallskip}\hline\noalign{\smallskip}                                  	                                  
Observed 				& $2$			& $0$	                & $11$	             \\
\noalign{\smallskip}\hline\noalign{\smallskip}                                  	                                  
Total SM 				& $4.1^{+2.6}_{-1.8}\,\,$ & $ 1.6^{+1.6}_{-1.1} \,\,$ & $ 4.2^{+3.4}_{-1.6}\,\,$ \\
     \noalign{\smallskip}\hline\noalign{\smallskip}           
$VV$ 			& $ 4.0  \pm  1.8 $ & $ 1.6  \pm  1.1 $ & $ 1.7  \pm  1.0 $ \\
Top          		& $ 0.15  \pm  0.11 $ & $ 0.04  \pm  0.03 $ & $ 0.8  \pm  0.4 $ \\
FNP 			& $0.0^{+0.2}_{-0.0}\,$   	& $0.0^{+0.1}_{-0.0}\,$ 	& $0.7^{+1.8}_{-0.7}\,$    \\
$\Zjet$                 & $0.0^{+1.8}_{-0.0}\,$  		& $0.0^{+1.2}_{-0.0}\,$        & $1.0^{+2.7}_{-1.0}\,$	  \\
Other &  --  &  --  &  --  \\
\noalign{\smallskip}\hline\noalign{\smallskip}
\end{tabular*}
}

\end{table}

\begin{table}[t]
\centering
  \caption{Background-only fits for SR3-WZ-0Ja to SR3-WZ-0Jc and SR3-WZ-1Ja to SR3-WZ-1Jc in the 3$\ell$ channel. All systematic and statistical uncertainties are included in the fit.
  }
  \label{tab:SR3LResultsWZ}
    \setlength{\tabcolsep}{0.0pc}
    {\small
      \begin{tabular*}{\textwidth}{@{\extracolsep{\fill}}lcccccc}
        \noalign{\smallskip}\hline\noalign{\smallskip}
        SR3-           & WZ-0Ja            & WZ-0Jb            & WZ-0Jc            & WZ-1Ja            & WZ-1Jb            & WZ-1Jc              \\[-0.05cm]
        \noalign{\smallskip}\hline\noalign{\smallskip}
\noalign{\smallskip}\hline\noalign{\smallskip}
Observed           & $21$              & $1$              & $2$              & $1$              & $3$              & $4$                    \\
\noalign{\smallskip}\hline\noalign{\smallskip}                                                                                                                                                                    
Total SM               & $21.7 \pm 2.9\;\,$     &    $ 2.7  \pm  0.5 $ & $ 1.56  \pm  0.33 $ &         $ 2.2  \pm  0.5 $ &  $1.82 \pm 0.26$          & $1.26 \pm 0.34$              \\
\noalign{\smallskip}\hline\noalign{\smallskip}
$WZ$           &  $ 19.5  \pm  2.9\;\,$ & $ 2.5  \pm  0.5 $ & $ 1.33  \pm  0.31 $ & $ 1.8  \pm  0.5 $ & $ 1.49   \pm  0.22 $ & $ 0.92  \pm  0.28 $ \\
$ZZ$          & $ 0.81  \pm  0.23 $ & $ 0.06  \pm  0.03 $ & $ 0.05  \pm  0.01 $ & $ 0.05  \pm  0.02 $ & $ 0.02  \pm  0.01 $ & --       \\
        $VVV$          & $0.31 \pm 0.07$       & $0.13 \pm 0.04$       & $0.13 \pm 0.03$        & $0.11 \pm 0.02$  & $0.12 \pm 0.03$          & $0.23 \pm 0.05$              \\
        $t\bar{t}V$    &$0.04 \pm 0.02$        & $0.01 \pm 0.01$       & $0.01 \pm 0.01$        & $0.14 \pm 0.04$  & $0.12 \pm 0.02$          & $0.08 \pm 0.02$              \\
        Higgs          & --                    & --                    &--                      &  $0.01 \pm 0.00$ & --                       &--                            \\
        FNP            & $1.1  \pm  0.5 $    & $0.02 \pm 0.01$       & $0.04 \pm 0.02$        & $0.11 \pm 0.06$  & $0.07 \pm 0.04$          & $0.01 \pm 0.00$              \\

        \noalign{\smallskip}\hline\noalign{\smallskip}
      \end{tabular*}
    }

\end{table}

\begin{table}[t]
\centering
  \caption{
Background-only fits for SR3-slep-a to SR3-slep-e in the 3$\ell$ channel. All systematic and statistical uncertainties are included in the fit.
  }
  \label{tab:SR3LResultsSlep}
    \setlength{\tabcolsep}{0.0pc}
    {\small
      \begin{tabular*}{\textwidth}{@{\extracolsep{\fill}}lccccc}
        \noalign{\smallskip}\hline\noalign{\smallskip}
        SR3-           & slep-a            & slep-b            & slep-c            & slep-d            & slep-e              \\[-0.05cm]
        \noalign{\smallskip}\hline\noalign{\smallskip}

Observed           & $4$              & $3$              & $9$              & $0$              & $0$                    \\
\noalign{\smallskip}\hline\noalign{\smallskip}

Total SM          & $ 2.2  \pm  0.8 $ & $ 2.8  \pm  0.4 $ & $ 5.4  \pm  0.9 $ & $ 1.4  \pm  0.4 $ & $ 1.14  \pm  0.23 $        \\
\noalign{\smallskip}\hline\noalign{\smallskip}
$WZ$          & $ 1.1  \pm  0.4 $ & $ 1.98  \pm  0.31 $ & $ 3.9  \pm  0.7 $ & $ 0.91  \pm  0.26 $ & $ 0.76  \pm  0.17 $       \\
$ZZ$          &$0.02 \pm 0.01$          & $0.01 \pm 0.01$          & $0.13 \pm 0.03$          & $0.06 \pm 0.02$          & $0.03 \pm 0.01$              \\
$VVV$          & $0.26 \pm 0.08$          & $0.34 \pm 0.05$          & $0.72 \pm 0.12$          & $0.36 \pm 0.10$          & $0.25 \pm 0.05$              \\
$t\bar{t}V$          & $0.07 \pm 0.03$          & $0.09 \pm 0.02$          & $0.20 \pm 0.04$          & $0.07 \pm 0.02$          & $0.02 \pm 0.01$              \\
Higgs          &  $0.01 \pm 0.00$          & $0.01 \pm 0.01$          & $0.03 \pm 0.02$          & $0.01 \pm 0.00$     & --              \\
FNP          & $0.80 \pm 0.46$          & $0.36 \pm 0.18$          & $0.48 \pm 0.25$          &    --      & $0.08 \pm 0.04$              \\
        \noalign{\smallskip}\hline\noalign{\smallskip}
      \end{tabular*}
    }

\end{table}

\begin{figure}[htb!]
\centering
\begin{subfigure}[t]{0.48\textwidth}\includegraphics[width=\textwidth]{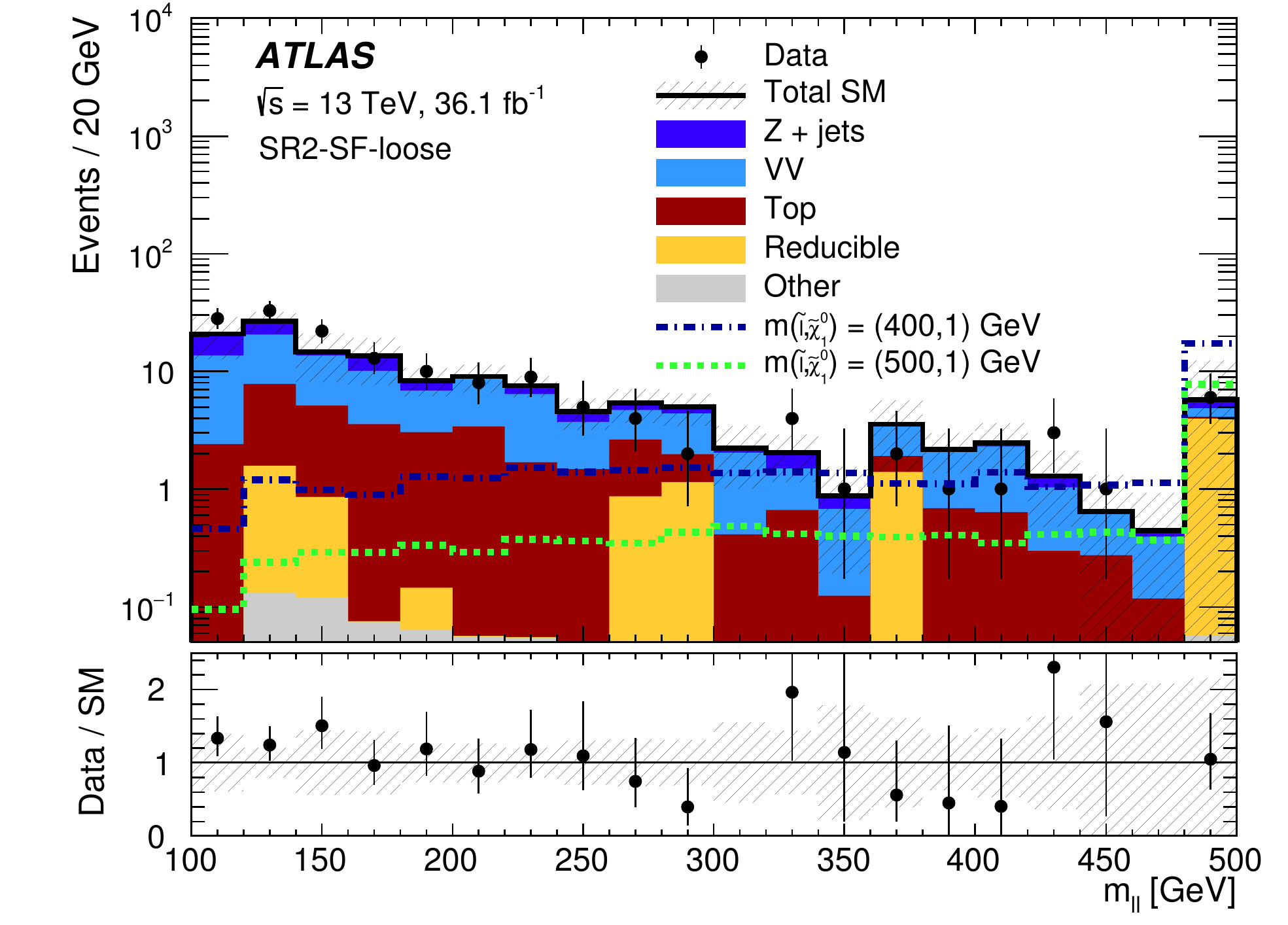}        
\caption{}\label{fig:SRloosemll}\end{subfigure}
\begin{subfigure}[t]{0.48\textwidth}\includegraphics[width=\textwidth]{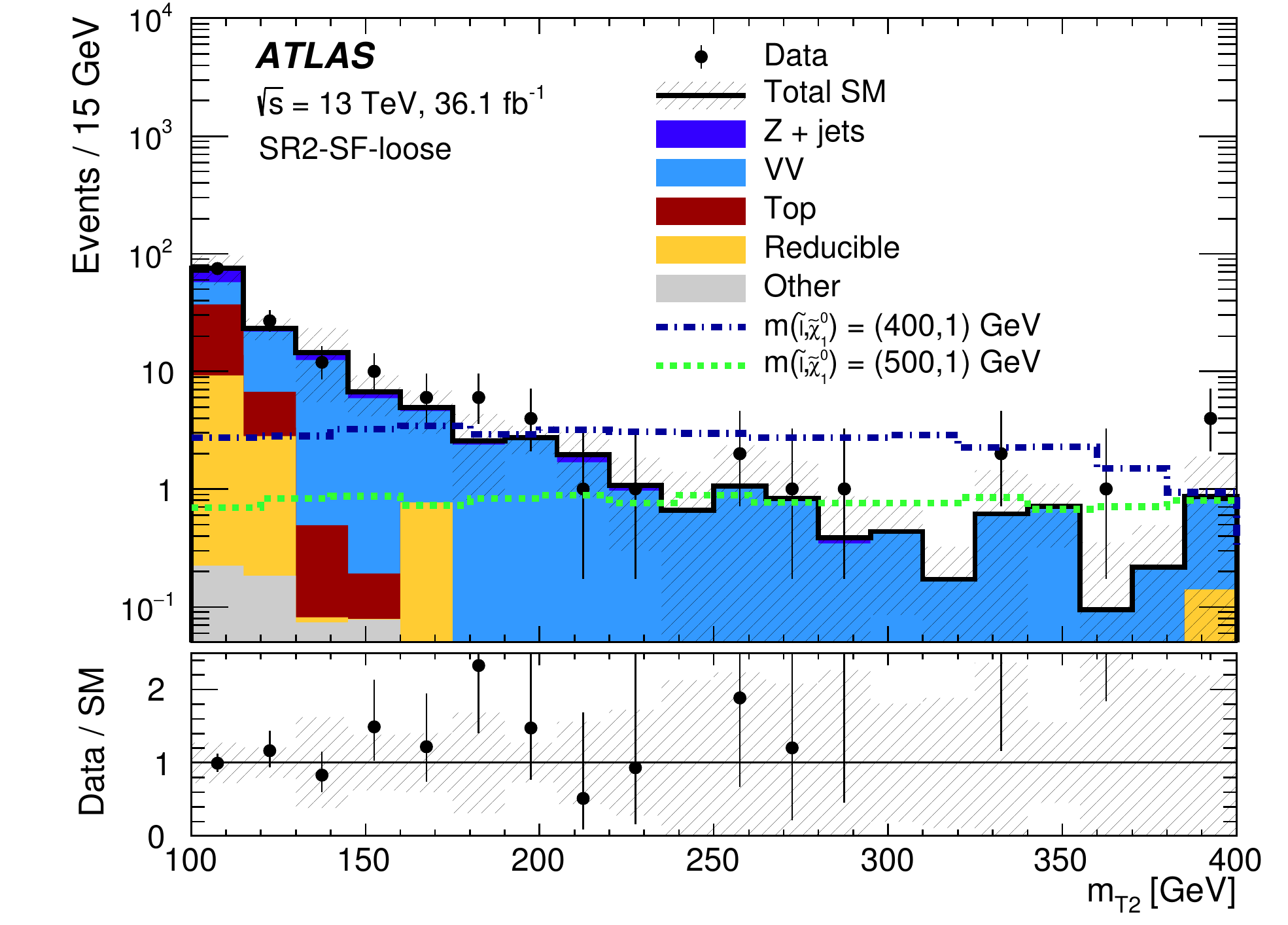} 
\caption{}\label{fig:SRloosemt2}\end{subfigure} \\
\begin{subfigure}[t]{0.48\textwidth}\includegraphics[width=\textwidth]{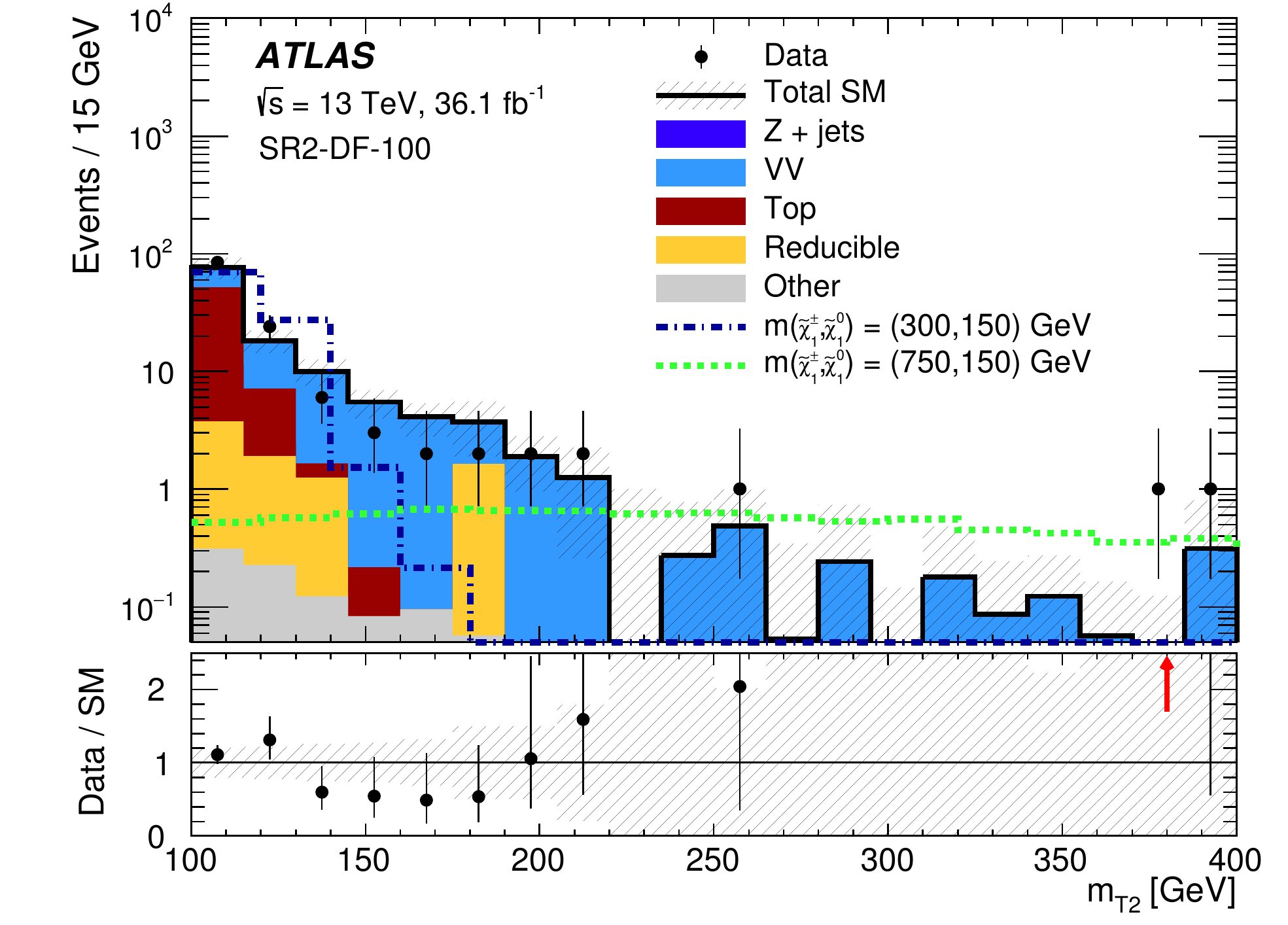} 
\caption{}\label{fig:2L0JDFmt2}\end{subfigure}
\caption{The (a) $m_{\ell\ell}$ and (b) \mttwo\ distributions for data and the estimated SM backgrounds in the 2$\ell$+0jets channel for SR2-SF-loose and (c) the \mttwo~distribution for the SR2-DF-100 selection. The normalization factors extracted from the corresponding CRs are used to rescale the $t\bar{t}$ and $VV$ contributions. The ``top'' background includes \ttbar\ and $Wt$, and the ``other'' backgrounds include Higgs bosons, $t\bar{t}V$ and $VVV$. The "reducible" category corresponds to the data-driven matrix method's estimate. The uncertainty bands include all systematic and statistical contributions. Simulated signal models for sleptons (a,b) or charginos (c) pair production are overlayed for comparison. The final bin in each histogram also contains the events in the overflow bin. The vertical red arrows indicate bins where the ratio of data to SM background, minus the uncertainty on this quantity, is larger than the $y$-axis maximum. }
\label{fig:Plot_SR2L0jets} 
\end{figure}

\begin{figure}[htbp]
\centering
\begin{subfigure}[t]{0.48\textwidth}\includegraphics[width=\textwidth]{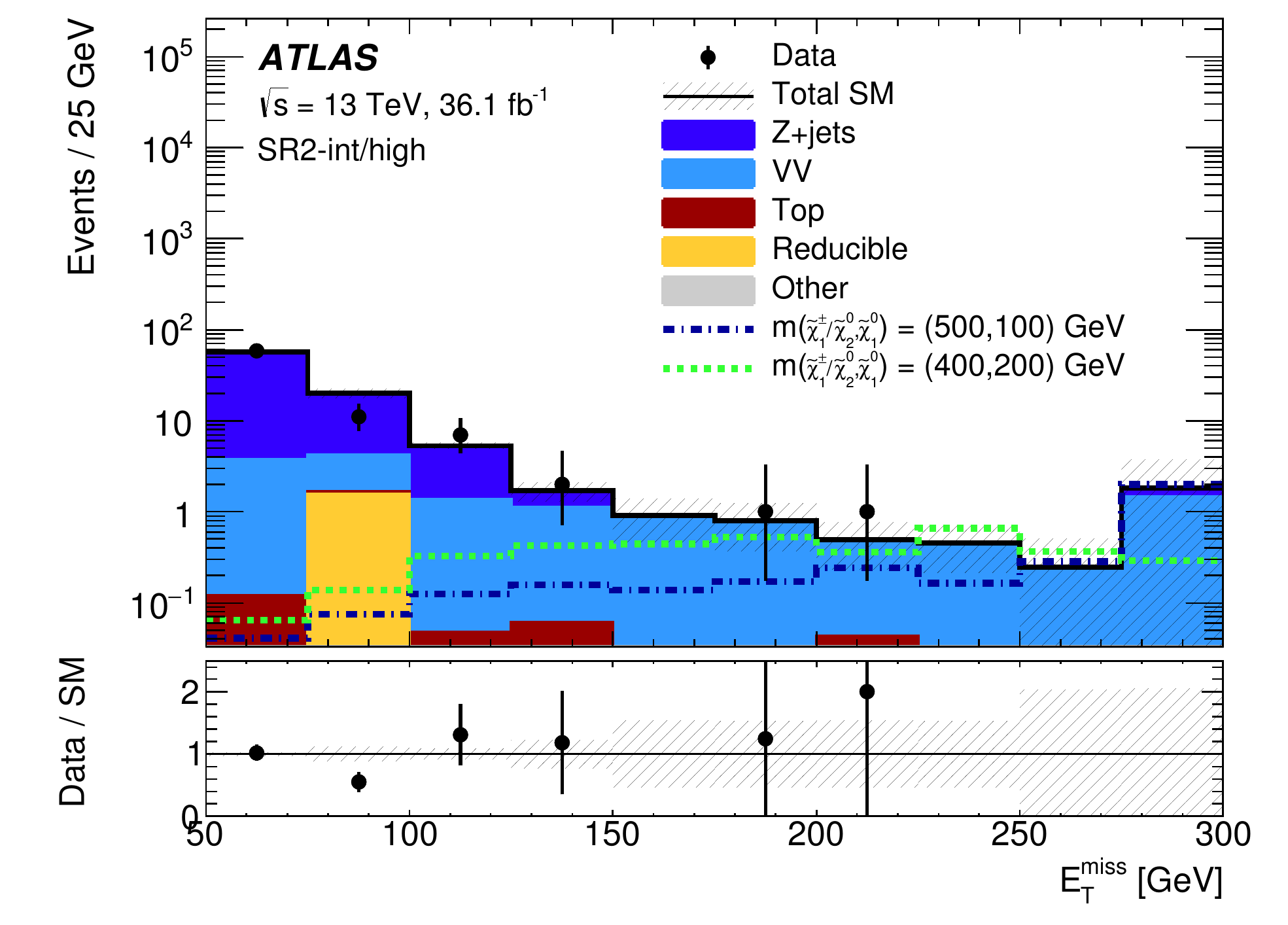}
\caption{}\label{fig:2L2JSRmed}\end{subfigure}
\begin{subfigure}[t]{0.48\textwidth}\includegraphics[width=\textwidth]{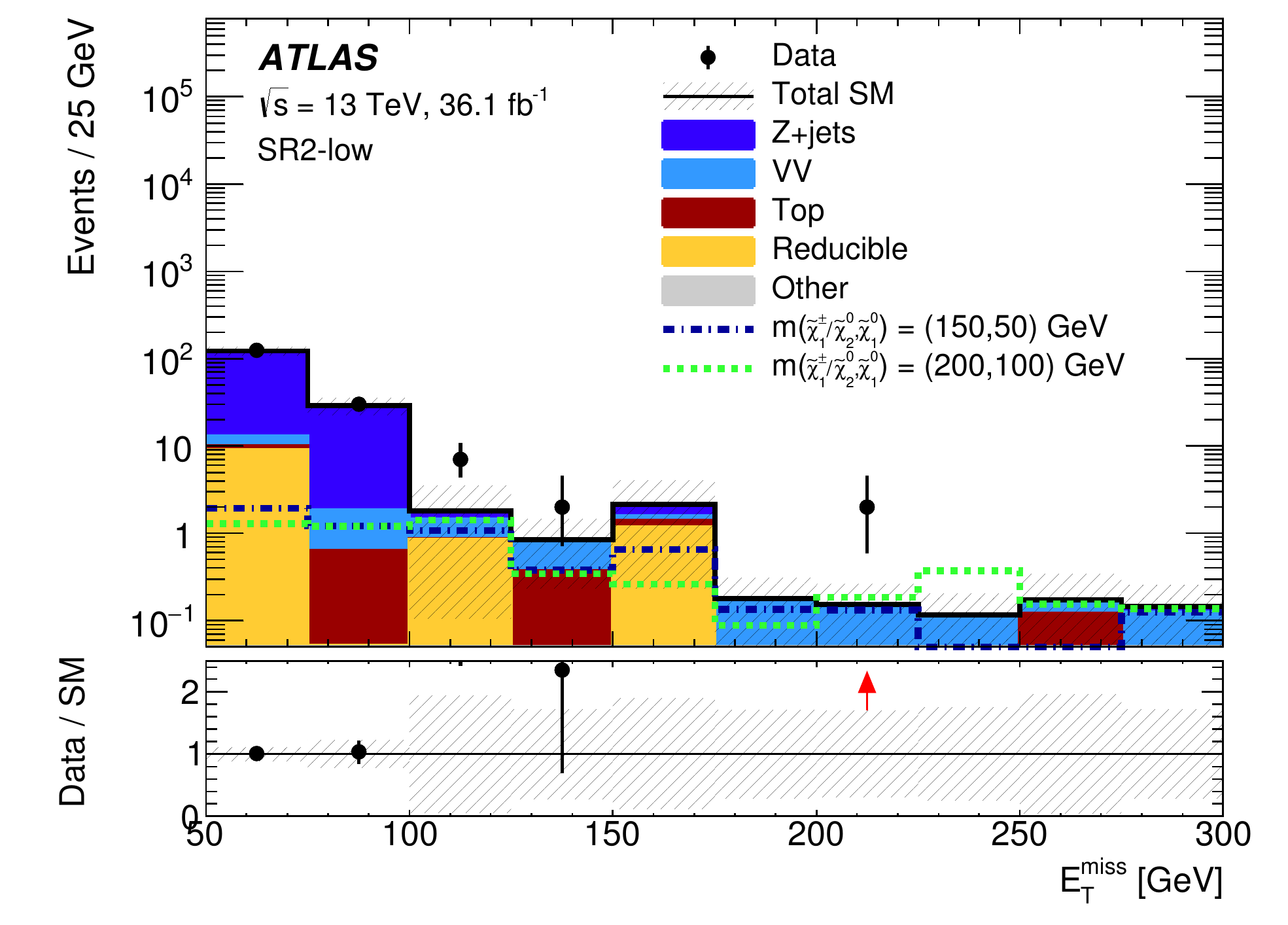}
\caption{}\label{2L2JSRlow}\end{subfigure} \\
\caption{\label{fig:Plot_SR2L2J} Distributions of \MET\ for data and the expected SM backgrounds in the 2$\ell$+jets channel for (a) SR2-int/high and (b) SR2-low, without the final $\MET$ requirement applied. The ``top'' background includes \ttbar\ , $Wt$ and $t\bar{t}V$, and the ``other'' backgrounds include Higgs bosons and $VVV$. The $Z$+jets contribution is evaluated using the data-driven $\gamma$+jet template method and the ``reducible'' category corresponds to the data-driven matrix method's estimate. The uncertainty bands include all systematic and statistical contributions.
Simulated signal models for charginos/neutralinos production are overlayed for comparison. The final bin in each histogram also contains the events in the overflow bin.
The vertical red arrows indicate bins where the ratio of data to SM background, minus the uncertainty on this quantity, is larger than the $y$-axis maximum. }
\end{figure}

\begin{figure}[htb!]
\centering
\begin{subfigure}[t]{0.48\textwidth}\includegraphics[width=\textwidth]{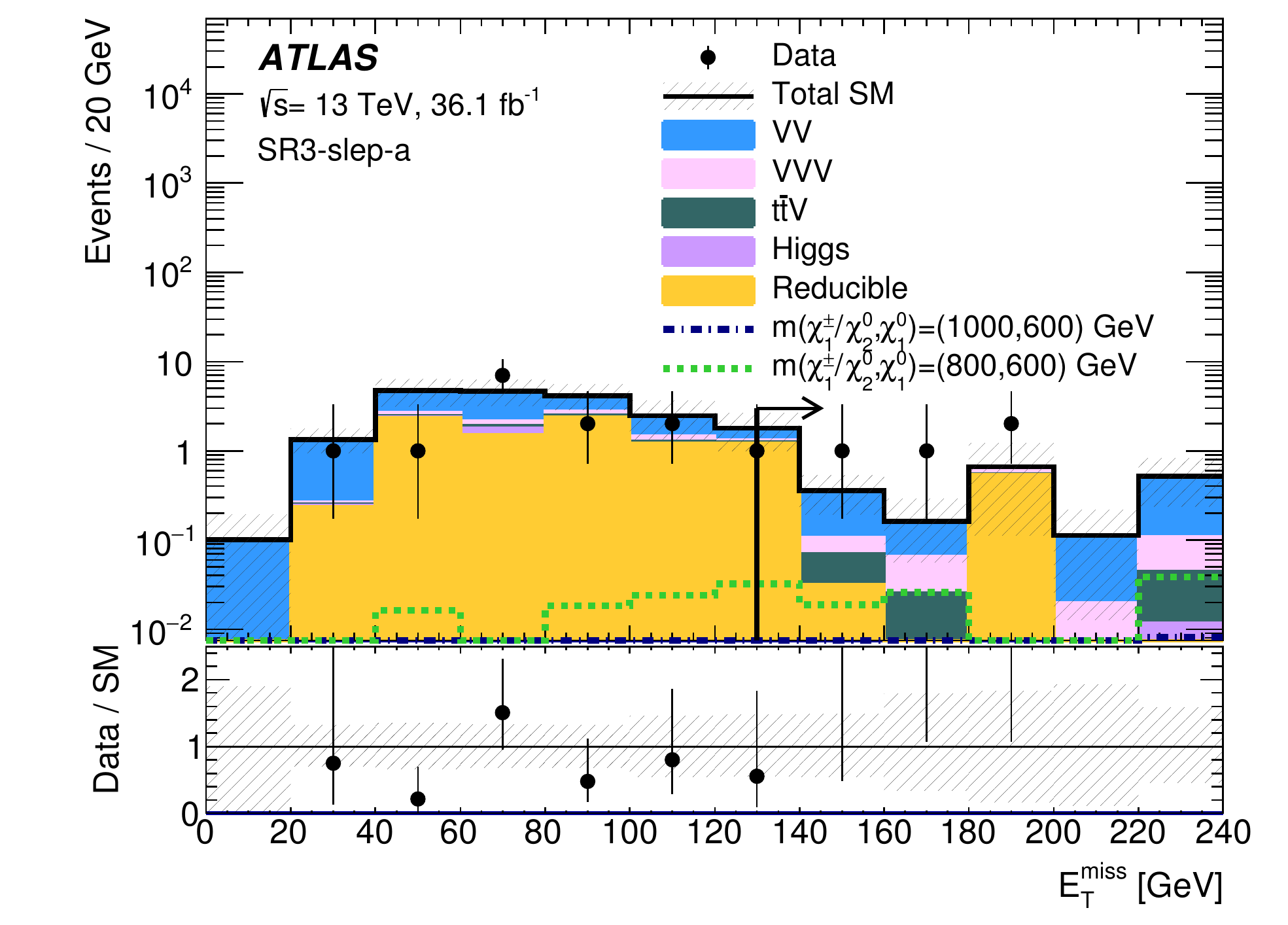}        
\caption{}\label{fig:3lSlepBinAMET}\end{subfigure}
\begin{subfigure}[t]{0.48\textwidth}\includegraphics[width=\textwidth]{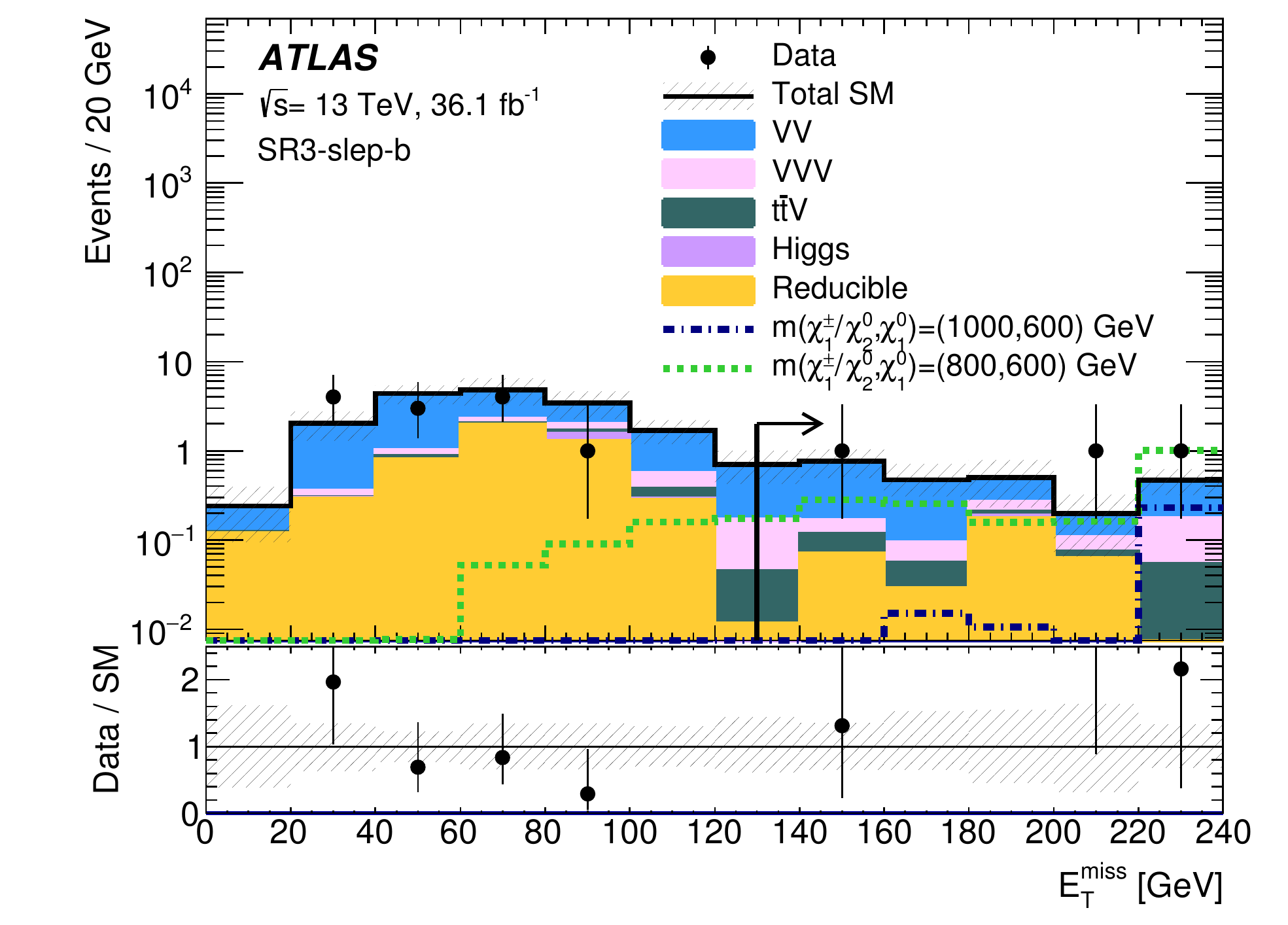}        
\caption{}\label{fig:3lSlepBinBMET}\end{subfigure}
\begin{subfigure}[t]{0.48\textwidth}\includegraphics[width=\textwidth]{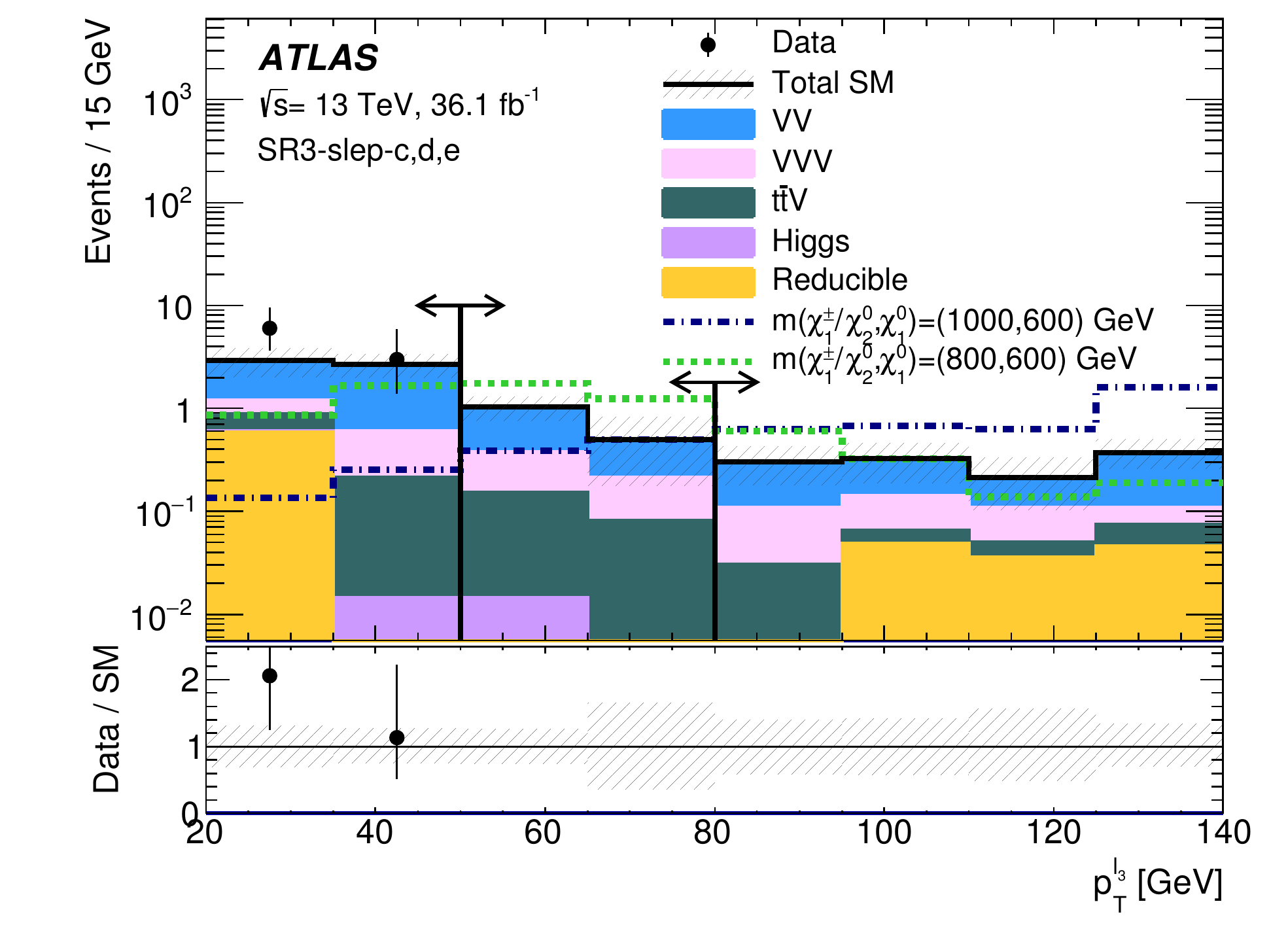}        
\caption{}\label{fig:3lSlepBinDEFpt}\end{subfigure}
\caption{Distributions of \MET\ for data and the estimated SM backgrounds in the 3$\ell$ channel for (a) SR3-slep-a and (b) SR3-slep-b and (c) distributions of the third leading lepton \pt\ in SR3-slep-c,d,e. The normalization factors extracted from the corresponding CRs are used to rescale the $WZ$ background. The ``reducible'' category corresponds to the data-driven fake-factor estimate. The uncertainty bands include all systematic and statistical contributions. Simulated signal models for charginos/neutralinos production are overlayed for comparison. The final bin in each histogram also contains the events in the overflow bin. }
\label{fig:Plot_3LSRslep} 
\end{figure}

\begin{figure}[htb!]
\centering
\begin{subfigure}[t]{0.48\textwidth}\includegraphics[width=\textwidth]{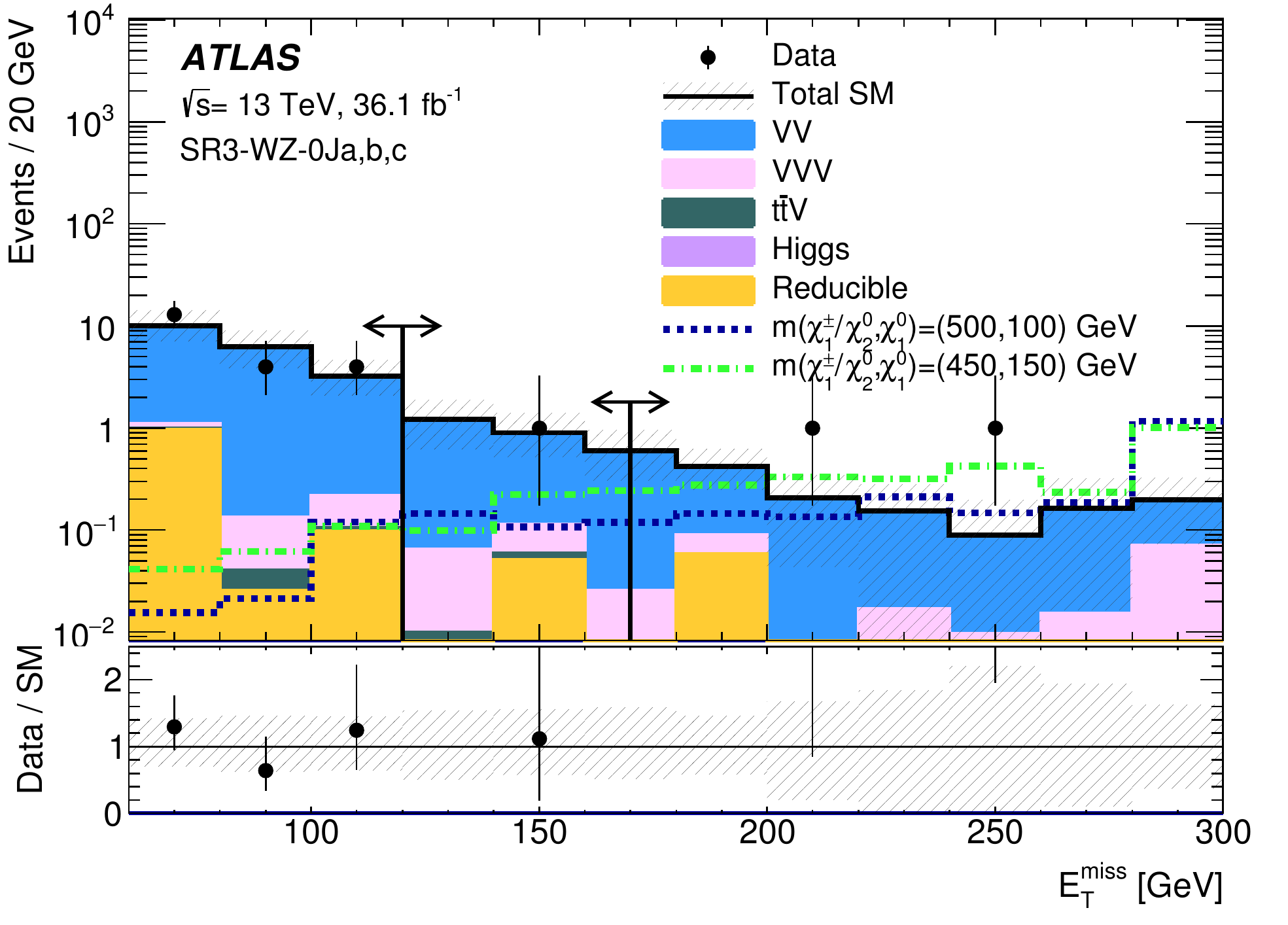}        
\caption{}\label{fig:3lWZBinABCMET}\end{subfigure}
\begin{subfigure}[t]{0.48\textwidth}\includegraphics[width=\textwidth]{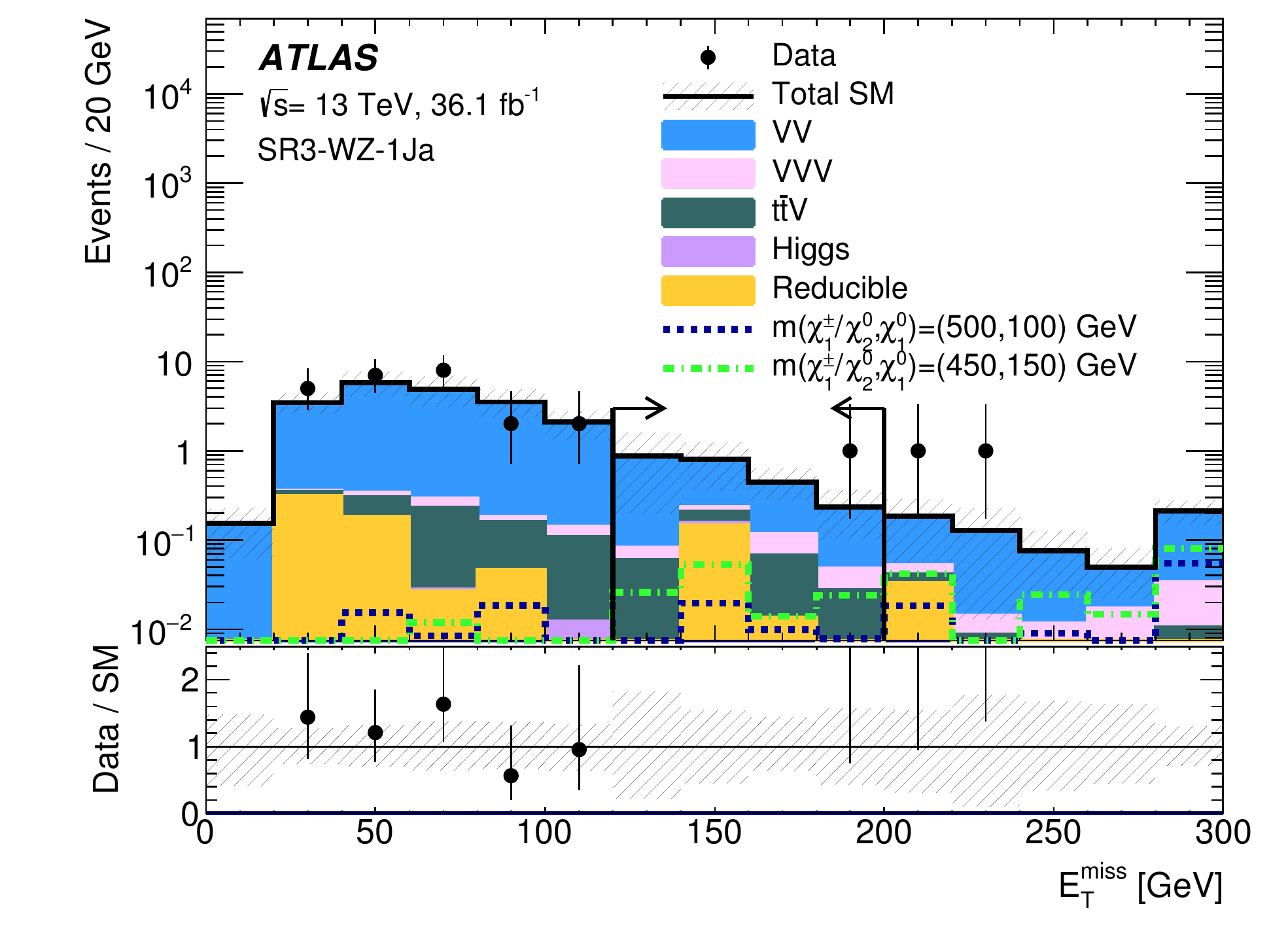}        
\caption{}\label{fig:3lSlepBinBMET}\end{subfigure}
\begin{subfigure}[t]{0.48\textwidth}\includegraphics[width=\textwidth]{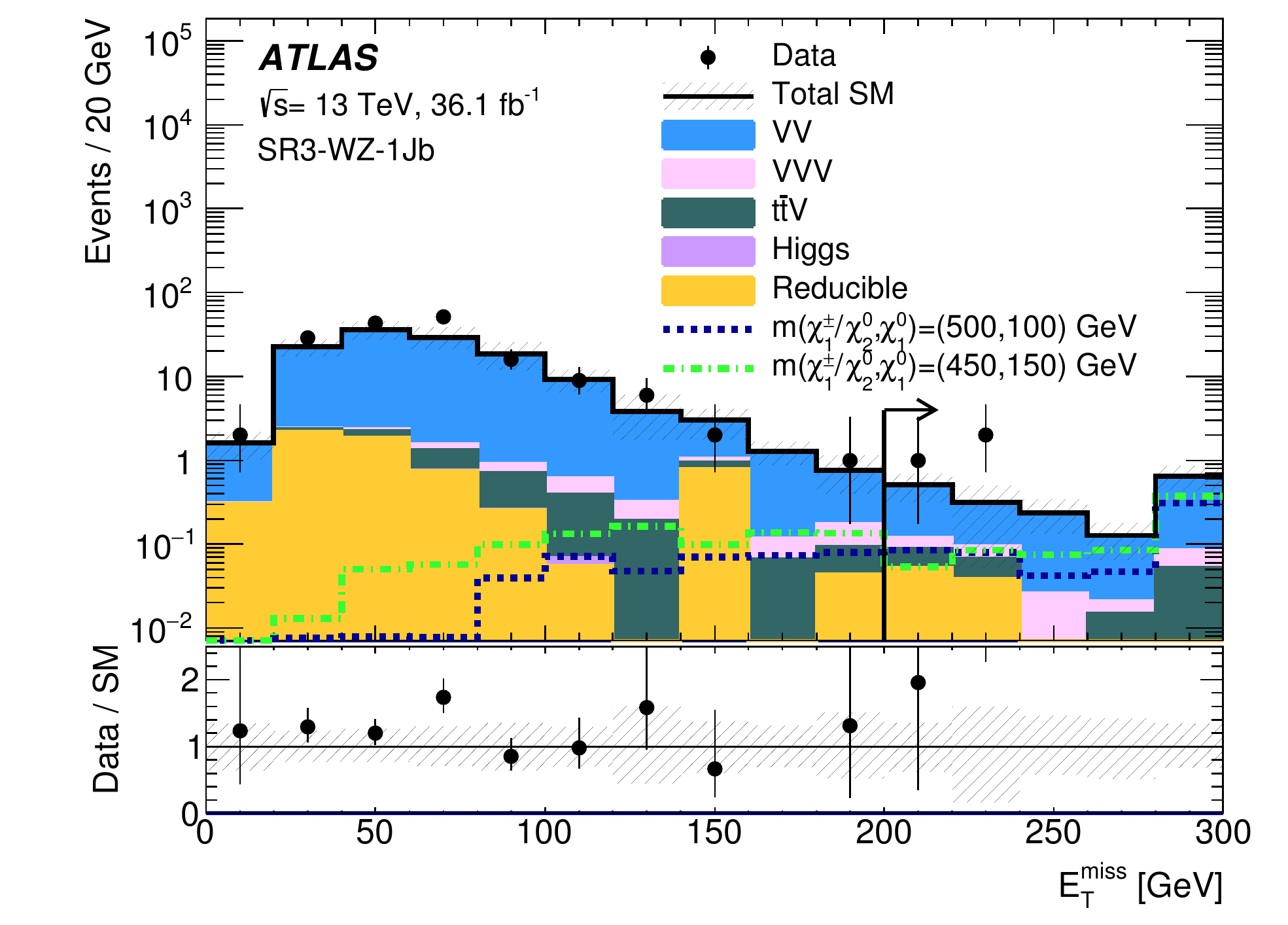}        
\caption{}\label{fig:3lSlepBinDEFpt}\end{subfigure}
\begin{subfigure}[t]{0.48\textwidth}\includegraphics[width=\textwidth]{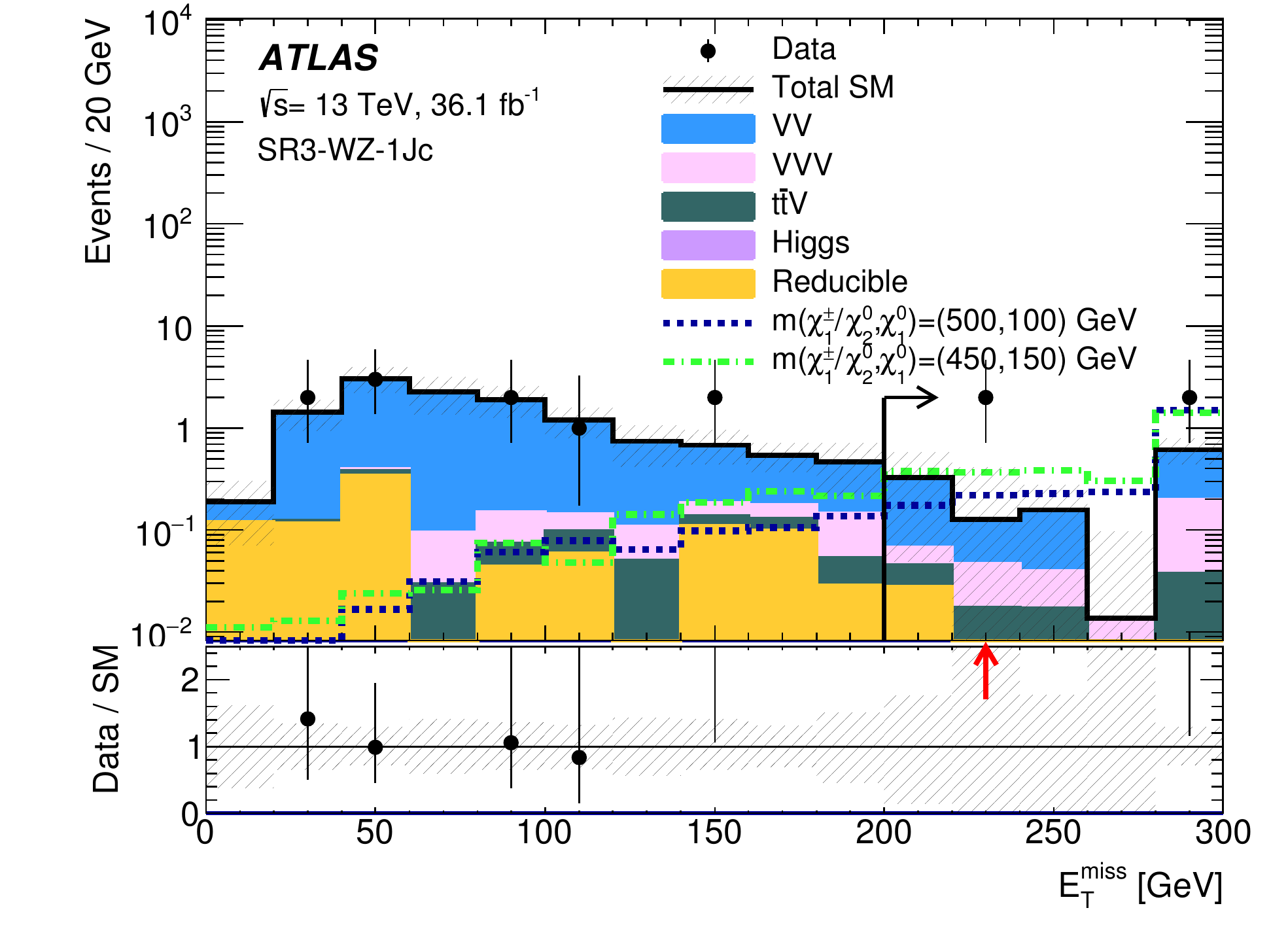}        
\caption{}\label{fig:3lSlepBinDEFpt}\end{subfigure}

\caption{Distributions of \MET\ for data and the estimated SM backgrounds in the 3$\ell$ channel for (a) SR3-WZ-0Ja,b,c,  (b) SR3-WZ-1Ja,  (c) SR3-WZ-1Jb and (d) SR3-WZ-1Jc. The normalization factors extracted from the corresponding CRs are used to rescale the 0-jet and $\geq$ 1-jet $WZ$ background components. The ``reducible'' category corresponds to the data-driven fake-factor estimate. The uncertainty bands include all systematic and statistical contributions. Simulated signal models for charginos/neutralinos production are overlayed for comparison. The final bin in each histogram also contains the events in the overflow bin.
The vertical red arrows indicate bins where the ratio of data to SM background, minus the uncertainty on this quantity, is larger than the $y$-axis maximum.}
\label{fig:Plot_3LSRWZ} 
\end{figure}

Figure~\ref{fig:Plot_SR2L0jets} shows a selection of kinematic distributions for data and the estimated SM backgrounds with their associated statistical and systematic uncertainties for the loosest inclusive SRs in the 2$\ell$+0jets channel: SR2-SF-loose and SR2-DF-100. The normalization factors extracted from the corresponding CRs are propagated to the $VV$ and $t\bar{t}$ contributions. 
Figure~\ref{fig:Plot_SR2L2J} shows the \MET\ distribution in SR2-int and SR2-high, which differ only in the \MET requirement, and in SR2-low of the 2$\ell$+jets channel. In the 3$\ell$ channel, distributions of \MET\ and the third leading lepton \pt\ are shown for the SR bins targeting $\slepton$-mediated decays in Figure~\ref{fig:Plot_3LSRslep} while Figure~\ref{fig:Plot_3LSRWZ} shows distributions of \MET\ in the bins targeting gauge-boson-mediated decays. Good agreement between data and expectations 
is observed in all distributions within the uncertainties.

\begin{figure}[htb!]
\captionsetup[subfigure]{justification=centering}
\centering
\begin{subfigure}[t]{0.48\textwidth}\includegraphics[width=\textwidth]{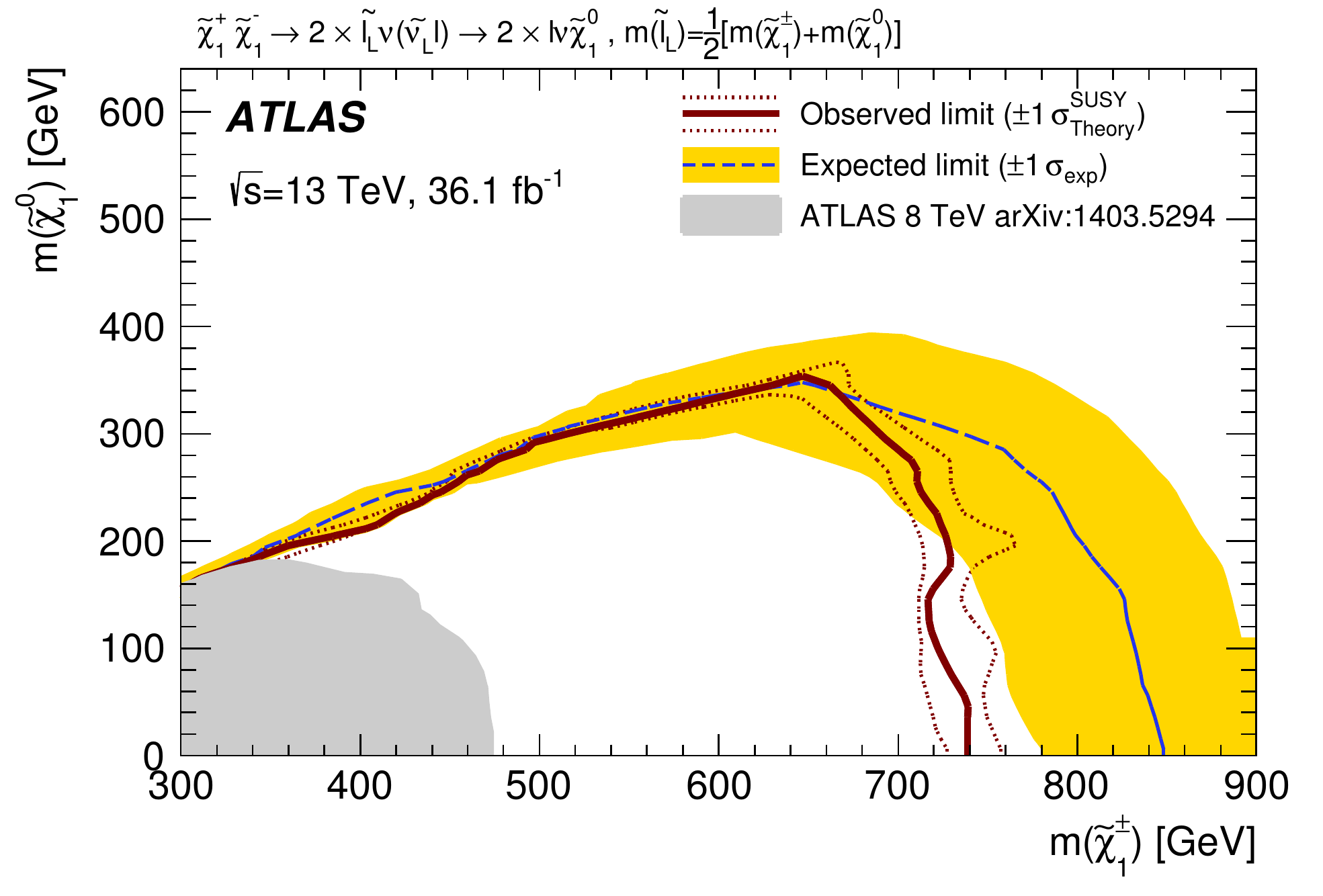}
\caption{\label{figa} 2$\ell$+0jets channel:\\ $\chinoonep\chinoonem$ production with $\slepton$-mediated decays}\label{fig:limits_C1C1slep}\end{subfigure} 
\begin{subfigure}[t]{0.48\textwidth}\includegraphics[width=\textwidth]{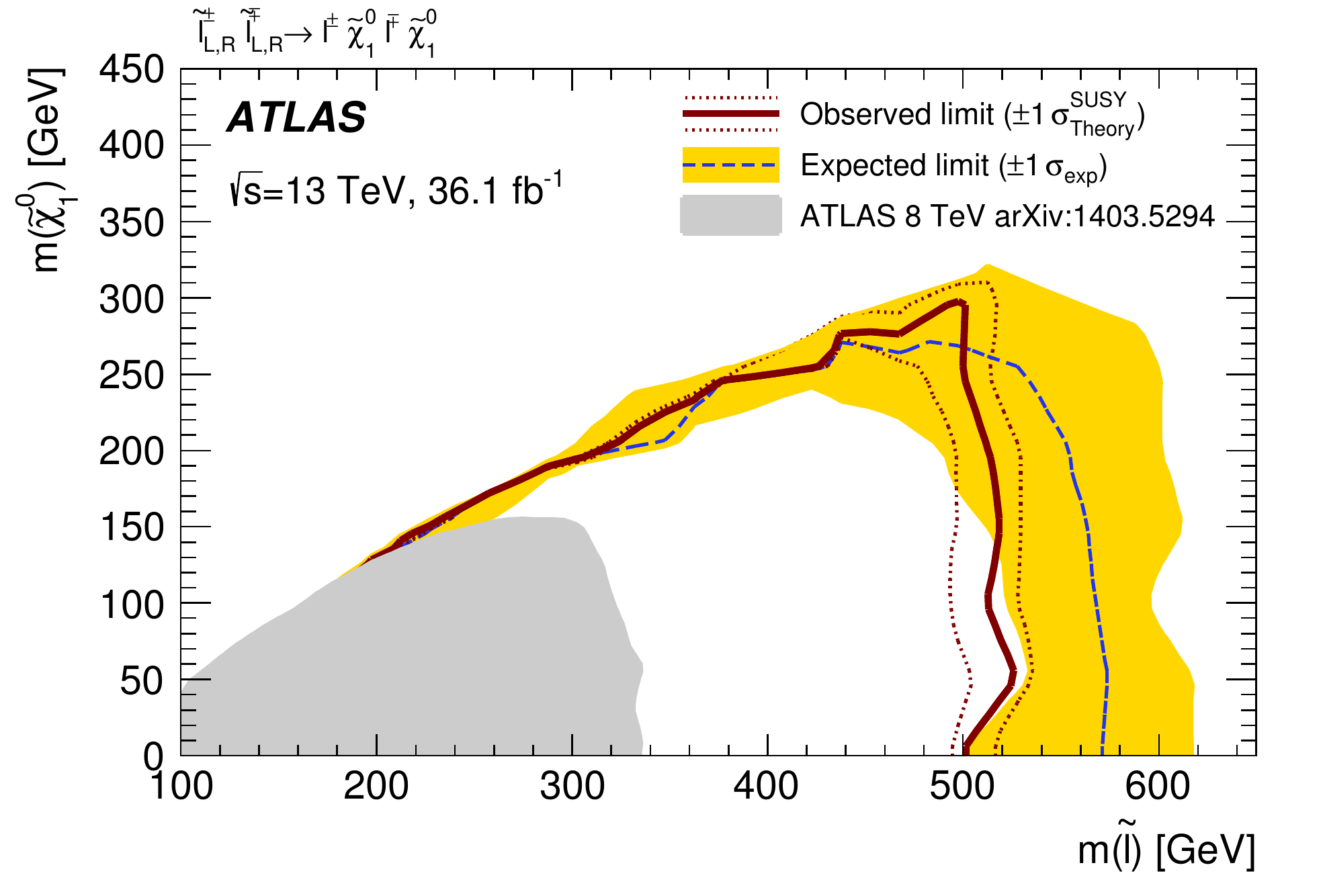}
\caption{\label{figb} 2$\ell$+0jets channel:\\ $\slepton\slepton$ production}\label{fig:limits_directslepton}\end{subfigure}
\begin{subfigure}[t]{0.48\textwidth}\includegraphics[width=\textwidth]{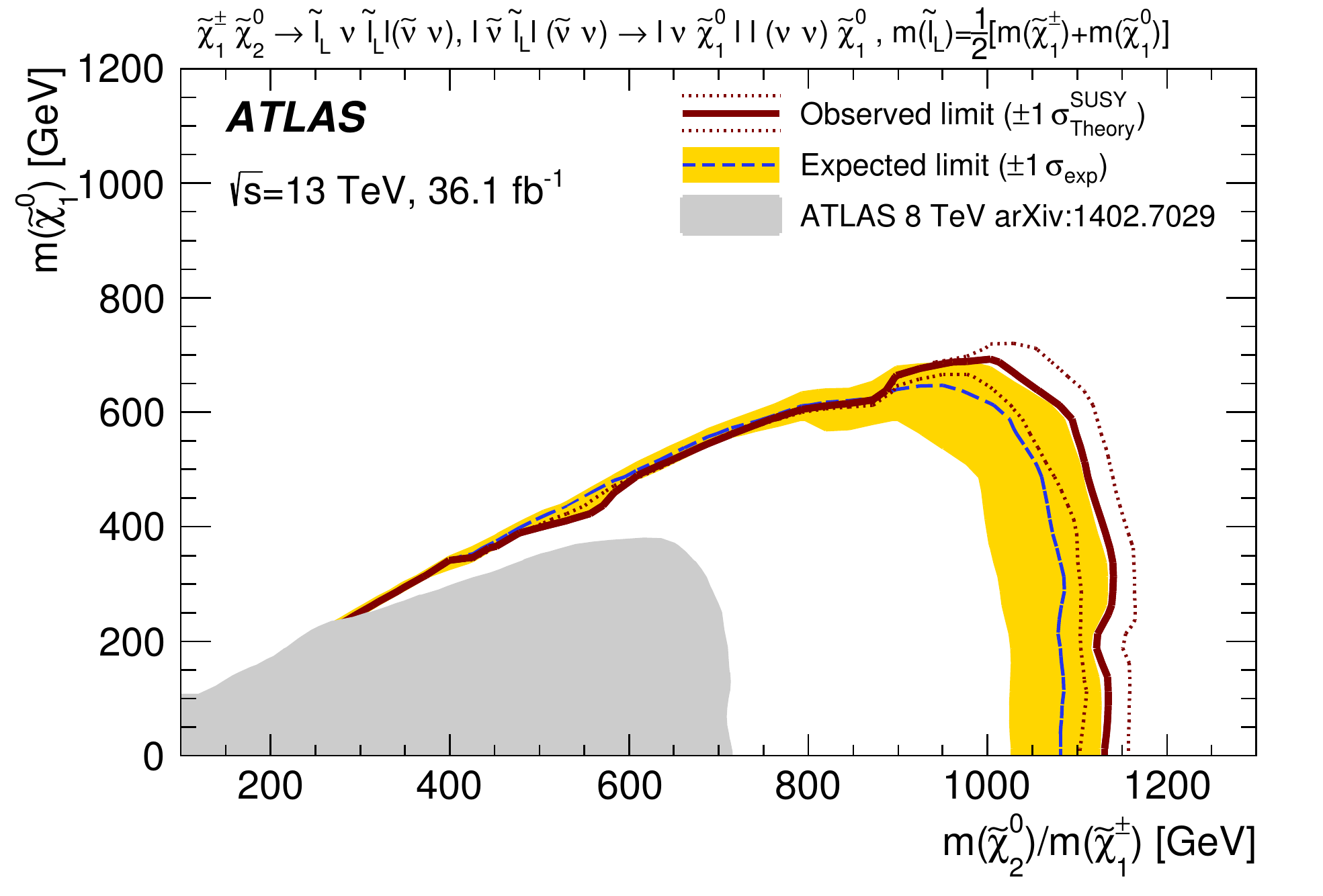}        
\caption{\label{figc} 3$\ell$ channel:\\ $\chinoonepm\ninotwo$ production with $\slepton$-mediated decays}\label{fig:limits_C1N2slep}\end{subfigure}
\begin{subfigure}[t]{0.48\textwidth}\includegraphics[width=\textwidth]{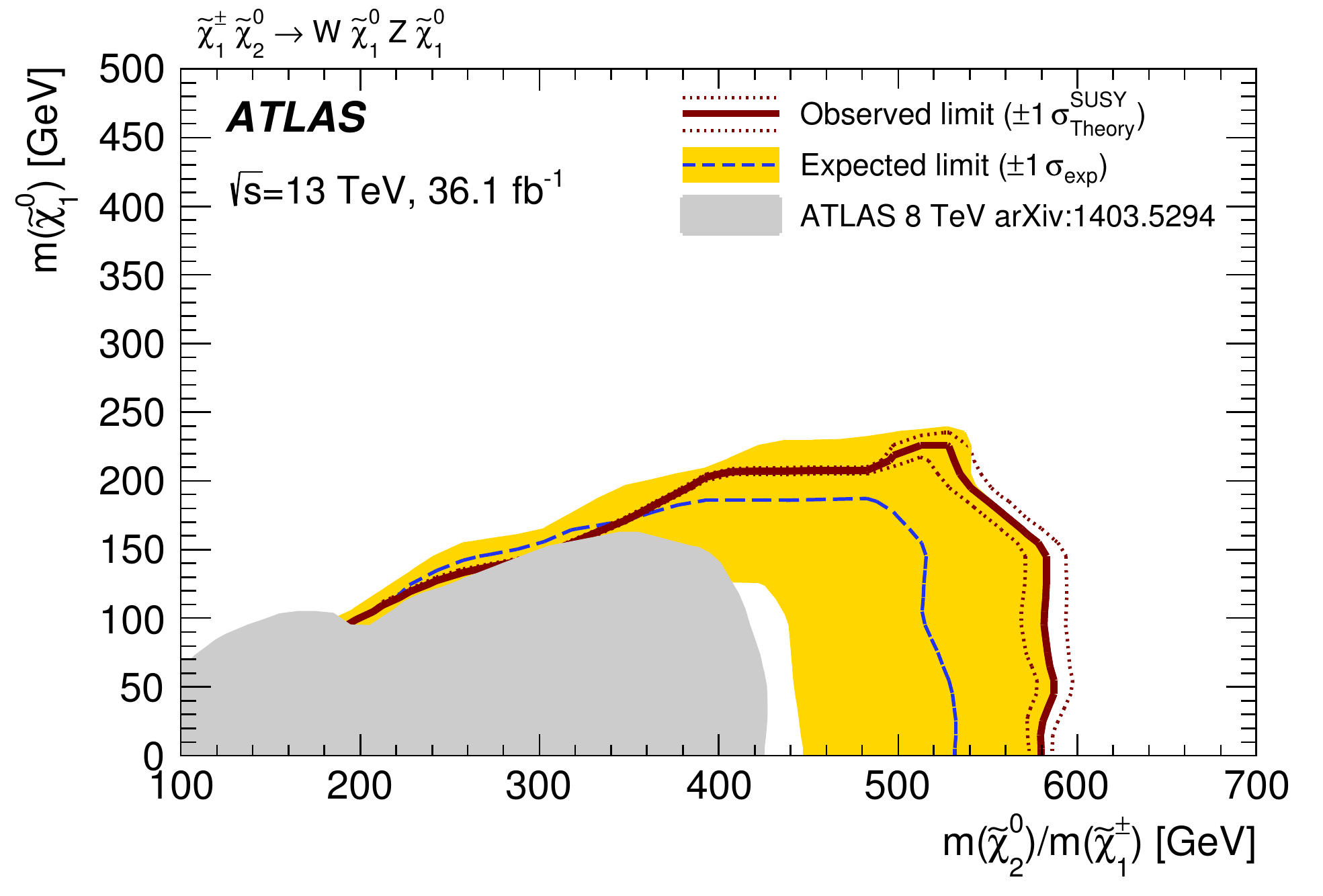}
\caption{\label{figd} 2$\ell$+jets and 3$\ell$ channels:\\ $\chinoonepm\ninotwo$ production with decays via $W/Z$}\end{subfigure}
\caption{
Observed and expected exclusion limits on SUSY simplified models for (a) chargino-pair production, (b) slepton-pair production, (c) chargino--neutralino production with slepton-mediated decays,
and (d) chargino--neutralino production with decays via $W/Z$ bosons. 
The observed (solid thick red line) and expected (thin dashed blue line) exclusion contours are indicated.
The shaded band corresponds to the $\pm$1$\sigma$ variations in the expected limit, including all uncertainties except theoretical
uncertainties in the signal cross-section.
The dotted lines around the observed limit illustrate the change in the observed limit as the nominal signal cross-section is scaled up and down by the theoretical uncertainty.
All limits are computed at 95\% confidence level. The observed limits obtained from ATLAS in Run 1 are also shown~\cite{SUSY-2013-11}.}
\label{fig:Results_LimitsAll} 
\end{figure}

In the absence of any significant excess, two types of exclusion limits for new physics scenarios are calculated using the CL$_\mathrm{s}$ prescription~\cite{Read:2002hq}. First, exclusion limits are set on the masses of the charginos, neutralinos, and sleptons for the simplified models in Figure~\ref{fig:TreeDiagramsPhysScenarios},
as shown in Figure~\ref{fig:Results_LimitsAll}.
Figures~\ref{figa} and \ref{figb} show the limits in the 2$\ell$+0jets channel in the models of direct chargino pair production with decays
via sleptons and direct slepton pair production, respectively.
Limits are calculated by statistically combining the mutually orthogonal exclusive SRs.
For the chargino pair model, all SF and DF bins are used and chargino masses up to 750 GeV are excluded at 95\% confidence level for a massless $\ninoone$ neutralino.
In the region with large chargino mass, the observed limit is weaker than expected because the data exceeds the expected backgrounds in SF-e, SF-f, and SF-g.
For the slepton pair model, which assumes mass-degenerate $\slepton_{\textup{L}}$ and $\slepton_{\textup{R}}$ states (where $\slepton=\tilde{e},\tilde{\mu},\tilde{\tau}$), only SF bins are used and slepton masses up to 500 GeV are excluded for a massless $\ninoone$ neutralino.

Figure~\ref{figc} shows the limits from the 3$\ell$ channel in the model of mass-degenerate chargino--neutralino pair production with decays via sleptons,
calculated using a statistical combination of the five SR3-slep regions. In this model, chargino and neutralino masses up to 1100~GeV are excluded for $\ninoone$ neutralino masses less than 550~GeV.

Figure~\ref{figd} shows the limits from the 3$\ell$ and 2$\ell$+jets channels in the model of mass-degenerate chargino--neutralino pair production with decays via $W/Z$ bosons.
The 3$\ell$ limits are calculated using a statistical combination of the six SR3-WZ regions.
Since the SRs in the 2$\ell$+jets channel are not mutually exclusive, the observed CL$_\mathrm{s}$ value is taken from the signal region with the best expected CL$_\mathrm{s}$ value.
The 3$\ell$ and 2$\ell$+jets channels are then combined, using the channel with the best expected CL$_\mathrm{s}$ value for each point in the model parameter space.
In this model, chargino and neutralino masses up to 580 GeV are excluded for a massless $\ninoone$ neutralino.

Second, model-independent upper limits are set on the visible signal cross-section ($\langle\epsilon{\rm \sigma}\rangle_{\rm{obs}}^{95}$) as well as on the observed ($S^{95}_{\rm{obs}}$) and expected ($S^{95}_{\rm{exp}}$) number of events from processes beyond-the-SM in the signal regions considered in this analysis.
The $p$-value and the corresponding significance for the background-only hypothesis are also evaluated.
For the 2$\ell$+0jets channel the inclusive signal regions defined in Table~\ref{tab:exclusiveSR} are considered whereas for the 3$\ell$ channel the calculation is performed for each bin separately. All the limits are at 95\% confidence level. The results can be found in Table~\ref{tab:upperlimits}.

\begin{table}[t]
\centering
\caption[Breakdown of upper limits.]{Summary of results and model-independent limits in the inclusive 2$\ell$+0jets, 2$\ell$+jets, and 3$\ell$ SRs.
The observed ($N_{\rm{obs}}$) and expected background ($N_{\rm{exp}}$) yields in the signal regions are indicated.
  Signal model-independent upper limits at 95\% confidence level on the the visible signal cross-section ($\langle\epsilon{\rm \sigma}\rangle_{\rm{obs}}^{95}$),
  and the observed and expected upper limit on the number of BSM events ($S^{95}_{\rm{obs}}$ and $S^{95}_{\rm{exp}}$, respectively) are also shown.
  The $\pm$1$\sigma$ variations of the expected limit originate from the statistical and systematic uncertainties in the background prediction.
  The last two columns show the $p$-value and the corresponding significance for the background-only hypothesis.
  For SRs where the data yield is smaller than expected, the $p$-value is truncated at 0.5 and the significance is set to 0.
\label{tab:upperlimits}}
\setlength{\tabcolsep}{0.0pc}
\begin{tabular*}{\textwidth}{@{\extracolsep{\fill}}llrccScSS}
\noalign{\smallskip}\hline\noalign{\smallskip}
{Signal channel}            & Region   & $N_{\rm{obs}}$ & $N_{\rm{exp}}$ & $\langle\epsilon{\rm \sigma}\rangle_{\rm{obs}}^{95}$[fb]  &  $S_{\rm{obs}}^{95}$  & 
$S_{\rm{exp}}^{95}$ & \text{\textit{p}(\textit{s}=0)} & $Z$  \\

\noalign{\smallskip}\hline\noalign{\smallskip}

2$\ell$+0jets & DF-100   &  $78$	 &     $68 \pm 7\;\,\,$ & $0.88$ &  32 & ${ 27}^{ +11 }_{  -8 }\;$ &  0.22 & 0.77 \\
              & DF-150   &  $11$ & $11.5 \pm 3.1\;\,\,$ & $0.32$ &  11.4 & $ { 12 }^{  +5 }_{  -4 }\;\,\,$ &  0.5  & 0    \\
                            
              & DF-200   &   $6$ &  $2.1\pm1.9$ & $0.33$ &  12.0 & $ { 10.3 }^{  +2.9 }_{  -1.9 }\;\,\,$ &  0.06 & 1.5 \\
              & DF-300   &  $ 2$ &  $0.6\pm0.6$ & $0.18$ &   6.6 & $ {  5.6 }^{  +1.1 }_{  -0.9 }$ &  0.10 & 1.3 \\
              & SF-loose & $153$ &   $133\pm22\;\,\,$ & $2.02$ &  73 & $ { 53 }^{ +21 }_{ -16 }$ &  0.16 & 1.0 \\
              & SF-tight &   $9$ &  $9.8\pm2.9$ & $0.29$ &  10.5 & $ { 12}^{  +4 }_{  -3 }\;\,\,$ &  0.5  & 0    \\

\noalign{\smallskip}\hline\noalign{\smallskip}

2$\ell$+jets &  SR2-int  &  $2$ & $4.1^{+2.6}_{-1.8}\;\,$ & $0.13$ &  4.5 & $ { 5.6 }^{ +2.2 }_{ -1.4 }$     &   0.5  & 0 \\
             &  SR2-high &  $0$ & $1.6^{+1.6}_{-1.1}\;\,$ & $0.09$ &  3.1 & $ { 3.1 }^{ +1.4 }_{ -0.1 }$     &   0.5  & 0 \\
             &  SR2-low  & $11$ & $4.2^{+3.4}_{-1.6}\;\,$  & $0.43$ &  15.7 & $ { 12 }^{ +4 }_{ -2 }\;\,\,$   &   0.06 & 1.6 \\

\noalign{\smallskip}\hline\noalign{\smallskip}

3$\ell$       &  WZ-0Ja & $21$  & $21.7 \pm 2.9\;\,$ & $0.35$ &  12.8 & $ { 14 }^{ +3 }_{ -5 }\;\,\,$ &   0.5  & 0     \\
              &  WZ-0Jb &  $1$ & $2.7 \pm 0.5$ & $0.10$ &  3.7 & $ { 4.6 }^{ +2.1 }_{ -0.9 }$   &   0.5  & 0     \\
              &  WZ-0Jc &  $2$ & $1.6 \pm 0.3$ & $0.13$ &  4.8 & $ { 4.1 }^{ +1.7 }_{ -0.7 }$  &  0.28 &0.57 \\
              &  WZ-1Ja &  $1$ & $2.2 \pm 0.5$ & $0.09$ &  3.2 & $ { 4.5 }^{ +1.6 }_{ -1.3 }$  &   0.5  & 0     \\
              &  WZ-1Jb & $ 3$ & $1.8 \pm 0.3$ & $0.16$ &  5.6 & $ { 4.3 }^{ +1.7 }_{ -0.9 }$ &  0.18 & 0.91 \\
              &  WZ-1Jc &  $4$ & $1.3 \pm 0.3$ &  $0.20$ &  7.2 & $ { 4.2 }^{ +1.7 }_{ -0.4 }$ & 0.03 & 1.8\\

              &  slep-a &  $4$ & $2.2 \pm 0.8$ & $0.19$ &  6.8 & $ { 4.7 }^{ +2.3 }_{ -0.5 }$ &  0.23 & 0.72  \\
              &  slep-b &  $3$ & $2.8 \pm 0.4$ & $0.14$ &  5.2 & $ { 5.1 }^{ +1.9 }_{ -1.2 }$ &  0.47 & 0.08 \\
              &  slep-c &  $9$ & $5.4 \pm 0.9$ & $0.29$ &  10.5 & $ { 6.8 }^{ +2.9 }_{ -1.3 }$ &  0.09 & 1.4\\
              &  slep-d &  $0$ & $1.4 \pm 0.4$  & $0.08$ &  3.0 & $ { 3.6 }^{ +1.2 }_{ -0.6 }$ &   0.5  & 0    \\
              &  slep-e &  $0$ & $1.1 \pm 0.2$  & $0.09$ &  3.3 & $ { 3.6 }^{ +1.3 }_{ -0.5 }$ &   0.5  & 0    \\
    
\noalign{\smallskip}\hline\noalign{\smallskip}
\end{tabular*}

\end{table}

\FloatBarrier

\section{Conclusion}
\label{sec:conclusion}
Searches for the electroweak production of neutralinos, charginos and sleptons decaying into final states with exactly two or three electrons or muons and missing transverse momentum are performed using 36.1 fb$^{-1}$ of $\sqrt{s}=13$~TeV proton--proton collisions recorded by the ATLAS detector at the Large Hadron Collider. Three different search channels are considered. The 2$\ell$+0jets channel targets direct $\chinoonep\chinoonem$ production where each $\chinoonepm$ decays via an intermediate $\slepton$, and direct $\slepton\slepton$ production. The 2$\ell$+jets channel targets associated $\chinoonepm\ninotwo$ production where each sparticle decays via an SM gauge boson giving a final state with two leptons consistent with a $Z$ boson and two jets consistent with a $W$ boson. Finally, the 3$\ell$ channel targets associated $\chinoonepm\ninotwo$ production with decays via either intermediate $\slepton$ or gauge bosons.

No significant excess above the SM expectation is observed in any of the signal regions considered across the three channels, and the results are used to calculate exclusion limits at 95\% confidence level in several simplified model scenarios. For associated $\chinoonepm\ninotwo$ production with $\slepton$-mediated decays, masses up to 1100~GeV are excluded for $\ninoone$ neutralino masses less than 550~GeV. Both the 2$\ell$+jets and 3$\ell$ channels place exclusion limits on associated $\chinoonepm\ninotwo$ production with gauge-boson-mediated decays. For a massless $\ninoone$ neutralino, $\chinoonepm$/$\ninotwo$ masses up to approximately 580~GeV are excluded. 
In the 2$\ell$+0jets channel, for direct $\chinoonep\chinoonem$ production with decays via 
an intermediate $\slepton$, masses up to 750 GeV are excluded for a massless $\ninoone$ neutralino.
For $\slepton\slepton$ production, masses up to 500 GeV are excluded for a massless $\ninoone$ neutralino, assuming mass-degenerate $\slepton_{\textup{L}}$ and $\slepton_{\textup{R}}$ (where $\slepton=\tilde{e},\tilde{\mu},\tilde{\tau}$).
These results significantly improve upon previous exclusion limits based on Run 1 data.
\section*{Acknowledgements}

We thank CERN for the very successful operation of the LHC, as well as the
support staff from our institutions without whom ATLAS could not be
operated efficiently.

We acknowledge the support of ANPCyT, Argentina; YerPhI, Armenia; ARC, Australia; BMWFW and FWF, Austria; ANAS, Azerbaijan; SSTC, Belarus; CNPq and FAPESP, Brazil; NSERC, NRC and CFI, Canada; CERN; CONICYT, Chile; CAS, MOST and NSFC, China; COLCIENCIAS, Colombia; MSMT CR, MPO CR and VSC CR, Czech Republic; DNRF and DNSRC, Denmark; IN2P3-CNRS, CEA-DRF/IRFU, France; SRNSFG, Georgia; BMBF, HGF, and MPG, Germany; GSRT, Greece; RGC, Hong Kong SAR, China; ISF and Benoziyo Center, Israel; INFN, Italy; MEXT and JSPS, Japan; CNRST, Morocco; NWO, Netherlands; RCN, Norway; MNiSW and NCN, Poland; FCT, Portugal; MNE/IFA, Romania; MES of Russia and NRC KI, Russian Federation; JINR; MESTD, Serbia; MSSR, Slovakia; ARRS and MIZ\v{S}, Slovenia; DST/NRF, South Africa; MINECO, Spain; SRC and Wallenberg Foundation, Sweden; SERI, SNSF and Cantons of Bern and Geneva, Switzerland; MOST, Taiwan; TAEK, Turkey; STFC, United Kingdom; DOE and NSF, United States of America. In addition, individual groups and members have received support from BCKDF, the Canada Council, CANARIE, CRC, Compute Canada, FQRNT, and the Ontario Innovation Trust, Canada; EPLANET, ERC, ERDF, FP7, Horizon 2020 and Marie Sk{\l}odowska-Curie Actions, European Union; Investissements d'Avenir Labex and Idex, ANR, R{\'e}gion Auvergne and Fondation Partager le Savoir, France; DFG and AvH Foundation, Germany; Herakleitos, Thales and Aristeia programmes co-financed by EU-ESF and the Greek NSRF; BSF, GIF and Minerva, Israel; BRF, Norway; CERCA Programme Generalitat de Catalunya, Generalitat Valenciana, Spain; the Royal Society and Leverhulme Trust, United Kingdom.

The crucial computing support from all WLCG partners is acknowledged gratefully, in particular from CERN, the ATLAS Tier-1 facilities at TRIUMF (Canada), NDGF (Denmark, Norway, Sweden), CC-IN2P3 (France), KIT/GridKA (Germany), INFN-CNAF (Italy), NL-T1 (Netherlands), PIC (Spain), ASGC (Taiwan), RAL (UK) and BNL (USA), the Tier-2 facilities worldwide and large non-WLCG resource providers. Major contributors of computing resources are listed in Ref.~\cite{ATL-GEN-PUB-2016-002}.

\printbibliography

\clearpage 
 
\begin{flushleft}
{\Large The ATLAS Collaboration}

\bigskip

M.~Aaboud$^\textrm{\scriptsize 34d}$,    
G.~Aad$^\textrm{\scriptsize 99}$,    
B.~Abbott$^\textrm{\scriptsize 125}$,    
O.~Abdinov$^\textrm{\scriptsize 13,*}$,    
B.~Abeloos$^\textrm{\scriptsize 129}$,    
S.H.~Abidi$^\textrm{\scriptsize 165}$,    
O.S.~AbouZeid$^\textrm{\scriptsize 143}$,    
N.L.~Abraham$^\textrm{\scriptsize 153}$,    
H.~Abramowicz$^\textrm{\scriptsize 159}$,    
H.~Abreu$^\textrm{\scriptsize 158}$
% ...
\bigskip
\\

% ...
$^{13}$Institute of Physics, Azerbaijan Academy of Sciences, Baku; Azerbaijan.\\
% ...
$^{34}$$^{(a)}$Facult\'e des Sciences Ain Chock, R\'eseau Universitaire de Physique des Hautes Energies - Universit\'e Hassan II, Casablanca;$^{(b)}$Centre National de l'Energie des Sciences Techniques Nucleaires (CNESTEN), Rabat;$^{(c)}$Facult\'e des Sciences Semlalia, Universit\'e Cadi Ayyad, LPHEA-Marrakech;$^{(d)}$Facult\'e des Sciences, Universit\'e Mohamed Premier and LPTPM, Oujda;$^{(e)}$Facult\'e des sciences, Universit\'e Mohammed V, Rabat; Morocco.\\
% ...
$^{99}$CPPM, Aix-Marseille Universit\'e, CNRS/IN2P3, Marseille; France.\\
% ...
$^{125}$Homer L. Dodge Department of Physics and Astronomy, University of Oklahoma, Norman OK; United States of America.\\
% ...
$^{129}$LAL, Universit\'e Paris-Sud, CNRS/IN2P3, Universit\'e Paris-Saclay, Orsay; France.\\
% ...
$^{143}$Santa Cruz Institute for Particle Physics, University of California Santa Cruz, Santa Cruz CA; United States of America.\\
% ...
$^{153}$Department of Physics and Astronomy, University of Sussex, Brighton; United Kingdom.\\
% ...
$^{158}$Department of Physics, Technion, Israel Institute of Technology, Haifa; Israel.\\
$^{159}$Raymond and Beverly Sackler School of Physics and Astronomy, Tel Aviv University, Tel Aviv; Israel.\\
% ...
$^{165}$Department of Physics, University of Toronto, Toronto ON; Canada.\\
% ...
$^{a}$ Also at  Department of Physics, University of Malaya, Kuala Lumpur; Malaysia.\\
$^{b}$ Also at Borough of Manhattan Community College, City University of New York, NY; United States of America.\\
$^{c}$ Also at Centre for High Performance Computing, CSIR Campus, Rosebank, Cape Town; South Africa.\\
$^{d}$ Also at CERN, Geneva; Switzerland.\\
$^{e}$ Also at CPPM, Aix-Marseille Universit\'e, CNRS/IN2P3, Marseille; France.\\
% ...
$^{*}$ Deceased

\end{flushleft}


\end{document}